\documentclass[superscriptaddress,preprintnumbers,aps,nofootinbib]{revtex4-1}

\usepackage{scrextend}

\usepackage{epsfig}
\usepackage{amsmath}
\usepackage{mathtools}
\usepackage{amsfonts}
\usepackage{amssymb}
\usepackage{color}

\definecolor{navyblue}{rgb}{0.0, 0.0, 0.5}
\definecolor{royalblue}{rgb}{0.25, 0.41, 0.88}
\definecolor{cadmiumgreen}{rgb}{0.0, 0.42, 0.24}
\definecolor{blue-violet}{rgb}{0.54, 0.17, 0.89}
\definecolor{darkviolet}{rgb}{0.58, 0.0, 0.83}
\definecolor{orange(colorwheel)}{rgb}{1.0, 0.5, 0.0}

\usepackage{hyperref}
\hypersetup{
    colorlinks=true, 
    linkcolor=royalblue, 
    citecolor=magenta}

\usepackage{booktabs}
\usepackage{multirow}
\usepackage{dcolumn}
\usepackage{colortbl}

\usepackage{stackengine}

\begin{document}

\title{Efficient Compression of Redshift-Space Distortion Data for Late-Time Modified Gravity Models}


\author{Yo Toda}
\email{y-toda@particle.sci.hokudai.ac.jp}
\affiliation{Department of Physics, Hokkaido University, Sapporo 060-0810, Japan}

\author{Adrià Gómez-Valent}
\email{agomezvalent@icc.ub.edu}
\affiliation{Departament de Física Quàntica i Astrofísica and Institut de Ciències del Cosmos, Universitat de Barcelona, Barcelona 08028, Catalonia, Spain}

\author{Kazuya Koyama}
\email{kazuya.koyama@port.ac.uk}
\affiliation{Institute of Cosmology and Gravitation, University of Portsmouth, Dennis Sciama Building, Portsmouth PO1 3FX, United Kingdom}

\preprint{EPHOU-24-011}

\begin{abstract}
Current cosmological observations allow for deviations from the standard growth of large-scale structures in the universe. These deviations could indicate modifications to General Relativity on cosmological scales or suggest the dynamical nature of dark energy. It is important to characterize these departures in a model-independent manner to understand their significance objectively and explore their fundamental causes more generically across a wider spectrum of theories and models. In this paper, we compress the information from redshift-space distortion data into 2-3 parameters $\mu_i$, which control the ratio between the effective gravitational coupling in Poisson's equation and Newton's constant in several redshift bins in the late universe. We test the efficiency of this compression using mock final-year data from the Dark Energy Spectroscopic Instrument (DESI) and considering three different models within the class of effective field theories of dark energy. The constraints on the parameters of these models, obtained from both the direct fit to the data and the projection of the compressed parameters onto the parameters of the models, are fully consistent, demonstrating the method's good performance. \footnote{Our code is available at \url{https://github.com/toda-cosmo/RSD}} Then, we apply it to current data and find hints of a suppressed matter growth in the universe at $\sim 2.7\sigma$ C.L., in full accordance with previous works in the literature. Finally, we perform a forecast with DESI data and show that the uncertainties on the parameters $\mu_1$ at $z<1$ and $\mu_2$ at $1<z<3$ are expected to decrease by approximately $40\%$ and $20\%$, respectively, compared to those obtained with current data. Additionally, we project these forecasted constraints onto the parameters of the aforesaid models. 

\end{abstract}

\maketitle

\section{Introduction}\label{sec:introduction}

The accelerated expansion of the universe is a very well-established observational fact, supported by a wide range of diverse cosmological measurements \cite{Planck:2018vyg,eBOSS:2020yzd,DES:2021wwk,Brout:2022vxf}. While in the context of the standard model of cosmology, also known as Lambda Cold Dark Matter ($\Lambda$CDM) model, this acceleration is explained by means of a cosmological constant $\Lambda>0$ with associated positive energy density $\rho_\Lambda=\Lambda/8\pi G$ and negative pressure $p_\Lambda=-\rho_\Lambda$ \citep{Peebles:2002gy,Padmanabhan:2002ji}, cosmological data still leaves room for dynamical forms of dark energy (DE) different from a rigid (immutable and homogeneous) $\Lambda$ \cite{Amendola:2015ksp,SolaPeracaula:2022hpd}. In fact, from a phenomenological perspective, deviations from the standard model might be necessary in view of the tensions afflicting it, such as the Hubble and growth tensions (see, e.g., the reviews \cite{Perivolaropoulos:2021jda,Abdalla:2022yfr} and references therein). They seem to require modifications to the background dynamics and/or the perturbation equations of $\Lambda$CDM.

In the last decade, these tensions have led to persistent hints of dynamical DE using cosmological data collected well before the advent of the Dark Energy Spectroscopic Instrument (DESI) \cite{Sahni:2014ooa,Sola:2015wwa, Sola:2016jky, Zhao:2017cud, SolaPeracaula:2017esw,SolaPeracaula:2018wwm, SolaPeracaula:2023swx,Gomez-Valent:2023uof,Gomez-Valent:2024tdb,Park:2024jns}. See, e.g., \cite{Poulin:2018cxd,SolaPeracaula:2019zsl,Jedamzik:2020krr,Seto:2021xua} for models introducing new physics before recombination as a means to alleviate the Hubble tension \footnote{DESI BAO data prefers a larger sound horizon $r_dh$ resulting in a higher Hubble constant when we consider early dark energy or a varying electron mass \cite{Qu:2024lpx,Seto:2024cgo}.}. The data on baryon acoustic oscillations from DESI's first-year data release, when combined with data on cosmic microwave background (CMB) from {\it Planck} \cite{Planck:2018vyg,Carron:2022eyg} and the Atacama Cosmology Telescope (ACT) \cite{ACT:2023kun}, as well as with the Type Ia supernovae (SNIa) from the Pantheon+ compilation \cite{Brout:2022vxf}, also point to a non-trivial evolution of the DE component. More concretely, using the Chevallier-Polarski-Linder ($w_0w_a$DE) parametrization of the DE equation of state (EoS) parameter \cite{Chevallier:2000qy, Linder:2002et}, $w(a)=w_{0}+(1-a)w_{a}$, the DESI collaboration finds $w_{0}=-0.831\pm0.066$
and $w_{a}=-0.73_{-0.28}^{+0.32}$ (68\% C.L.) \cite{DESI:2024mwx}. This suggests a hint of dynamical dark energy at a significance level of approximately 2.6$\sigma$. Remarkably, the evidence increases to 3.5$\sigma$ or 3.9$\sigma$ when the Pantheon+ compilation is replaced with the Union3 \cite{Rubin:2023ovl} or DES-SN5YR \cite{DES:2024hip} SNIa, respectively. Consistent results are also obtained with more flexible reconstruction techniques of the DE \cite{DESI:2024aqx}.

Data on large-scale structure (LSS) have played an important role in constraining cosmological models and may be relevant for consolidating the aforementioned hints of dynamical DE. The data on redshift-space distortions (RSD) and weak gravitational lensing tend to favor a lower growth of matter perturbations in the late-time universe compared to the {\it Planck} best-fit $\Lambda$CDM model \cite{Macaulay:2013swa,Joudaki:2017zdt,Gomez-Valent:2017idt,Nesseris:2017vor,Gomez-Valent:2018nib,Benisty:2020kdt,Wright:2020ppw,Nunes:2021ipq,Nguyen:2023fip,Adil:2023jtu}. Interestingly, upcoming RSD data from DESI will cover a redshift range up to $z\sim 2.1$, extending by $\sim 0.6$ units the range covered by past surveys, and with small relative uncertainties of $3-6\%$ in the intermediate range $z\in (0.4,1.6)$ \cite{DESI:2023dwi}. These new data will allow us to better assess the status of the growth tension.

In this work, we aim to compress the information contained in RSD data into variables that parameterize deviations from the standard growth of perturbations in the late universe, with the aid of a couple of sensible priors. These variables should have a crystal-clear physical interpretation and be applicable to a broad range of modified gravity and dark energy models. More concretely, in this paper, we model the gravitational coupling as a step function and introduce
2 or 3 parameters (depending on the case under study) representing the step widths of the functions. We
demonstrate that this model can reproduce the growth of density fluctuations
of effective field theories of DE (EFTofDE) with observational accuracy. Other models, such as Galileon
gravity, Dvali-Gabadadze-Porrati (DGP) gravity, and $f(R)$ gravity, have already been examined
in \cite{Denissenya:2017uuc}. We apply the compression technique proposed in \cite{Denissenya:2017uuc} and study its robustness and universality by applying it to three different models within the large class of EFTofDE. In most works in the literature, the time dependence of the EFT parameters or the effective gravitational coupling is assumed when deriving or forecasting constraints from observations \cite{Zhao:2008bn,Planck:2015bue,Ferte:2017bpf,DES:2018ufa,Sakr:2021ylx,DES:2022ccp,Casas:2022vik}. These constraints strongly depend on the assumed functional form. In \cite{Casas:2017eob}, Casas et al. introduced 6 bins at $0<z<3$ and performed forecasts for the constraints from CMB, galaxy clustering and weak lensing. Our focus in this work is galaxy clustering and we show that 2-3 bins are already enough to describe a wide variety of models. More general (continuous) reconstructions of the gravitational coupling can be performed using more sophisticated techniques, such as a Bayesian reconstruction or Gaussian Processes \cite{Raveri:2021dbu,Pogosian:2021mcs,Ruiz-Zapatero:2022xbv,Mu:2023zct}, or even Principal Component Analysis (PCA) \cite{Pogosian:2010tj,Hojjati:2011xd,Asaba:2013xql}, but the resulting constraints cannot be easily projected onto constraints on the parameters of specific models. The PCA approach uses a large number of bins to identify and extract the data's well-constrained modes, which are insensitive to the number of bins used in the analysis. In principle, a theory prediction can be mapped to these principal component modes. However, because the latter depends on data covariance, this projection is data-dependent. Of course there are models that will not be well described by only 2-3 bins. In this case, the use of more bins or alternative methods as PCA are complementary to our approach. See also \cite{Zheng:2023yco} for a very model-independent method to constrain deviations from standard growth that does not make use of RSD data.

The 2-3 parameter description proposed in \cite{Denissenya:2017uuc} offers a quite model-independent way to reveal the physical characteristics of the underlying theory required to explain the RSD data. It relies on the cosmological principle and the covariant conservation of matter. It can account, e.g., for phenomena such as the suppression of growth at late times \cite{Macaulay:2013swa,Joudaki:2017zdt,Gomez-Valent:2017idt,Nesseris:2017vor,Gomez-Valent:2018nib,Benisty:2020kdt,Wright:2020ppw,Nunes:2021ipq,Nguyen:2023fip,Adil:2023jtu}. Additionally, we will see that it provides an efficient method for constraining EFT parameters with specific time dependencies by projecting the compressed information onto the latter. 

The paper is organized as follows. In Sec. \ref{sec:binning_method} we explain our method and, in Sec. \ref{sec:EFT}, the basics of the EFTofDE and, in particular, the three models studied in this work. Sec. \ref{sec:data} is devoted to the description of the data sets. As a proof of concept, we test the method using mock data from DESI forecast and demonstrate that our compression conserves the statistical content of the original data, keeping the same level of constraining power. We achieve this by comparing the constraints on the EFT parameters obtained from a direct fit to the RSD data with those derived from fitting the compressed data. Both approaches yield essentially the same results. We present this analysis in Sec. \ref{sec:DESI}. In Sec. \ref{sec:VI} we apply the method to currently available data from various galaxy surveys and also perform a forecast with mock data from DESI. Finally, our conclusions are presented in Sec. \ref{sec:conclusions}.


\section{Binning strategy}\label{sec:binning_method}

We aim to translate the information contained in the RSD data into constraints on the effective gravitational coupling $G_{\rm eff}$ entering the Poisson equation and examine whether this can be achieved without any loss of statistical power. $G_{\rm eff}$ can encapsulate pure deviations from the Newton constant $G$ or other effects, such as the clustering of DE in non-standard cosmologies. We parameterize the departures from the usual Poisson equation in terms of the function $\mu$ as follows \cite{Pogosian:2010tj},

\begin{equation}
k^2\psi(k,a)= -4\pi G_{\rm eff}(a)a^2\sum_{i}\rho_i(a)\Delta_i(k,a) \qquad {\rm with}\qquad    G_{\rm eff}(a)= \mu(a) G\,,
\end{equation}
where $\rho_i$ and $\Delta_i$ are the background energy densities and the gauge-invariant density contrasts of the various species $i$, respectively, and $\psi$ is the scalar potential in the Newtonian gauge, see Appendix \ref{sec:AppendixA} for details. We consider a time- (or, equivalently, redshift- or scale-factor-) dependent function $\mu(a)$, and, therefore, we assume no significant dependence on the scale $k$. 

In this work, we also assume the covariant conservation of matter and the equivalence principle, apart from the cosmological principle at the background level. Under these conditions, at deep subhorizon scales, the evolution of matter perturbations during and after the matter-dominated era is governed by the equation of the matter density contrast $\delta_m= \delta\rho_m/\rho_m$,
\begin{equation}
\ddot{\delta}_m+2H\dot{\delta}_m-4\pi G\mu\rho_m\delta_m=0  \,,\label{eq:delta_evolve}
\end{equation}
with the dots denoting derivatives with respect to the cosmic time $t$ and $H=\dot{a}/a$ the Hubble function. 
 
We approximate the effective gravitational coupling as a step function, using the two binning strategies for $\mu$ already tested in \cite{Denissenya:2017uuc}. We call these parametrizations $\mu$-2param and $\mu$-3param, for obvious reasons:
\newline 
\newline
\underline{$\mu$-2param}

\begin{equation}\label{eq:2p}
\mu(a)=
    \begin{cases}
        \mu_1 & \text{if }\,a>0.5\quad (z<1),\, 
        \\
        \mu_2 & \text{if } \,0.25<a<0.5 \quad(1<z<3),\,
    \\
        1 & \text{if }\,a<0.25 \quad (z >3),\,
    \end{cases}
\end{equation}
\newline\newline
\underline{$\mu$-3param}

\begin{equation}\label{eq:3p}
\mu(a)=
    \begin{cases}
        \mu_1 & \text{if }\,a>0.5\quad (z<1),\,
        \\
        \mu_2 & \text{if } \,0.25<a<0.5 \quad(1<z<3),\, 
        \\
        \mu_3 & \text{if }\, 0.1<a<0.25 \quad (3< z <9),\, 
        \\
        1 & \text{if }\, a<0.1 \quad (z > 9),\, 
    \end{cases}
\end{equation}
The redshift ranges covered by these parametrizations extend beyond those of current and future RSD data, see Sec. \ref{sec:data}. The compression of the RSD data into the various $\mu_i$ is only meaningful if this compressed information is employed to constrain modified gravity theories that only allow for deviations from GR in the redshift range covered by the parametrization itself. The constraints obtained with $\mu$-2param can be in principle employed to constrain models with possible departures from $\mu=1$ at $z<3$, whereas $\mu$-3param can be used if these departures enter at $z<9$. This is the case for the models described in Sec. \ref{sec:EFT}. Earlier modifications of gravity would obviously require the introduction of more bins at higher redshifts. It is also important to bear in mind that these alternative models must also respect the basic assumptions listed above, such as the covariant conservation of matter.


\section{The Effective Field Theory of dark energy}\label{sec:EFT}
In this paper, we use the EFTofDE to test whether the 2-3 parameter description is accurate enough for stage-IV surveys such as DESI. The most general scalar-tensor theory of gravity that leads to second-order field equations is known as Horndeski theory \cite{Horndeski:1974wa,Deffayet:2011gz}. Its action reads, 

\begin{equation}
S=\int d^{4}x\sqrt{-g}\left[\sum_{i=2}^{5} 
\frac{\mathcal{L}_{i}}{8 \pi G}
+\mathcal{L}_{m}\right]\,,
\end{equation}
where $\mathcal{L}_m$ is the Lagrangian density of the matter sector, which includes the Standard Model of Particle Physics and possible extensions of the latter accounting for dark matter and the neutrino masses, and the $\mathcal{L}_i$'s with $i\in [2,5]$ describe the gravity sector, with

\begin{align}
\mathcal{L}_{2} & =K(\phi,X)\,,\nonumber \\
\mathcal{L}_{3} & =-G_{3}(\phi,X)\square\phi\,,\nonumber \\
\mathcal{L}_{4} & =G_{4}(\phi,X)R+G_{4,X}(\phi,X)\left[(\square\phi)^{2}-\phi_{;\mu\nu}\phi^{;\mu\nu}\right]\,,\nonumber \\
\mathcal{L}_{5} & =G_{5}(\phi,X)G_{\mu\nu}\phi^{;\mu\nu}-\frac{1}{6}G_{5,X}(\phi,X)\left[(\square\phi)^{3}+2\phi_{;\mu}^{\,\nu}\phi_{;\nu}^{\,\alpha}\phi_{;\alpha}^{\,\mu}-3\phi_{;\mu\nu}\phi^{;\mu\nu}\square\phi\right]\,.
\end{align}
The functions $K$ and the $G_i$'s can depend on the scalar field $\phi$ and its kinetic term $X=-\partial_\mu\phi\partial^\mu\phi/2$. Here the ``$;$'' denote covariant derivatives and $\square\phi\equiv \nabla^\mu\nabla_\mu\phi$. In the context of the EFTofDE it is shown that the linear perturbations in Horndeski theory are controlled by just four functions $\alpha_j$ ($j=M,K,B,T$), which are related to the Horndeski functions as follows \cite{Bellini:2014fua},

\begin{align}
HM_{*}^{2}\alpha_{M}\equiv & \,\frac{d}{dt}M_{*}^{2},\label{eq:am} \\
HM_{*}^{2}\alpha_{K}\equiv & \,2X\left(K_{,X}+2XK_{,XX}-2G_{3,\phi}-2XG_{3,\phi X}\right)+\cdot\cdot\cdot \\
HM_{*}^{2}\alpha_{B}\equiv & \,2\dot{\phi}\left(XG_{3,X}-G_{4,\phi}-2XG_{4,\phi X}\right)\nonumber \\
 & +8XH\left(G_{4,X}+2XG_{4,XX}-G_{5,\phi}-XG_{5,\phi X}\right)\nonumber \\
 & +2\dot{\phi}XH^{2}\left(3G_{5,X}+2XG_{5,XX}\right),\label{eq:ab} \\
 HM_{*}^{2}\alpha_{T}\equiv & \,2X\left[2G_{4,X}-2G_{5,\phi}-\left(\ddot{\phi}-\dot{\phi}H\right)G_{5,X}\right],
\end{align}
with $M_{*}^{2}\equiv2\left(G_{4}-2XG_{4,X}+XG_{5,X}-\dot{\phi}HXG_{5,X}\right)$. The function $\alpha_M$ controls the running of the effective Planck mass. The kineticity $\alpha_K$ affects the scalar perturbations' kinetic energy and, in particular,  the scalar sound speed. The kinetic braiding $\alpha_B$ describes the mixing of the scalar and metric kinetic terms, and controls the clustering of dark energy. Finally, $\alpha_T=c_T^2-1$ parametrizes deviations of the speed of propagation of gravitational waves from the speed of light. In the limit $\alpha_i\to 0$ we recover standard General Relativity (GR). For dedicated reviews on EFTofDE and Horndeski's theory we refer the reader to \cite{Kase:2018aps,Frusciante:2019xia}.

There exist very tight constraints on $\alpha_T$ at $z\sim 0$ obtained from the analysis of the gravitational wave (GW) event GW170817 and its electromagnetic counterpart \cite{LIGOScientific:2017vwq}, $\left|\alpha_{T}\right|\lesssim10^{-15}$ \cite{Baker:2017hug,Creminelli:2017sry,Sakstein:2017xjx,Ezquiaga:2017ekz}. It is possible to build attractor models within the Horndeski class with GWs propagating at the speed of light when $z\to 0$ but not in the past \cite{Amendola:2018ltt}. This would still respect the constraint of \cite{LIGOScientific:2017vwq}. However, in this study, for the sake of simplicity, we assume $\alpha_{T}=0$ $\forall{z}$ and also neglect $\alpha_{K}$, since it does not affect
constraints on other parameters at leading order at the level of linear perturbations as it does not affect the evolution of the density perturbation under the quasi-static approximation.  \cite{Bellini:2015xja,Alonso:2016suf}. 

Under these assumptions we have  $G_3=G_3(X)$, $G_4=G_4(\phi)$ and $G_5={\rm const.}$, so we can write 
\begin{align}
HM_{*}^{2}\alpha_{M} =& 2\dot{\phi}G_{4,\phi}\,, \\
HM_{*}^{2}\alpha_{B} =& 2\dot{\phi}\left(XG_{3,X}-G_{4,\phi}\right)\,.
\end{align}
Still, $\alpha_{M}$ and $\alpha_{B}$ are independent free functions, and their observational constraints depend greatly on their concrete forms. Under the quasi-static approximation, the modification to the effective Newton constant encapsulated in $\mu$ can be 
expressed in terms of $\alpha_{\mathrm{B}}$ and $\alpha_{\mathrm{M}}$ as 
\begin{equation}\label{eq:Geff}
  \mu = 1 + \frac{ 2 c_{\mathrm{sN}}^{2} \left(1- M_{*}^{2} \right) + \left( \alpha_{B} + 2 \alpha_{M} \right)^{2}}{2c_{\mathrm{sN}}^{2} M_{*}^{2}},
\end{equation}
where $c_{\text{sN}}^{2}= D c_{\text{s}}^{2}$, $D = \alpha_{K}+ \frac{3}{2} \alpha_{B}^2$, 
\begin{equation}
    c_{\text{s}}^{2}=  \frac{1}{D}\left[\left(2-\alpha_{B}\right)\left(-\frac{H^{\prime}}{aH^{2}}+\frac{1}{2}\alpha_{B}+\alpha_{M}\right)-\frac{8 \pi G \left(\rho_{\textrm{tot}}+p_{\textrm{tot}}\right)}{H^{2}M_{*}^{2}}+\frac{\alpha_{B}^{\prime}}{aH}\right] \label{eq:cs2}
\end{equation}
is the sound speed squared \cite{Pogosian:2016pwr} and $\rho_{\textrm{tot}}$ and $p_{\textrm{tot}}$ are the total density and pressure, respectively. Here the prime denotes the derivative with respect to the conformal time. 
The stability condition requires $D>0$ and $c_{\text{s}}^{2} \geq 0$. The quasi-static approximation works well below the sound horizon and we assume that this holds on scales relevant for the galaxy clustering observation. See Refs. \cite{Brando1, Brando2} for relativistic corrections in these models. 
We note that $\mu$ is independent of $\alpha_K$ as mentioned earlier under the quasi-static approximation. However, $\alpha_K$ still plays an important role for the stability condition.  

In this work we consider only late-time deviations from GR through the following three functional forms (with $i=\{B,M\}$),

\begin{itemize}
    \item propto-Omega:   $\alpha_i(a)=c_i\,\Omega_{\rm DE}(a)$,
    \item Inv-Hubble-Squared: $\alpha_i(a)=c_i\,\left[H_0/H(a)\right]^2$,
    \item propto-Scale: $\alpha_i(a)=c_i\,a$,
     
\end{itemize}
with $H_0$ the Hubble parameter and $\Omega_{\rm DE}(a)$ the DE fraction. In the propto-Omega model, we vary $c_B$ and $c_M$ separately, i.e., considering $c_B\ne 0$ with $c_M=0$ or $c_M\ne 0$ with $c_B=0$. This gives rise to two different sub-families of models. In the propto-Scale model, instead, we vary $c_B$ setting $c_M=0$ or $c_M=-c_B$\footnote{In the propto-Omega model, the the condition $\alpha_B=-\alpha_M$ triggers an instability in the radiation-dominated era \cite{Noller:2020afd}, so we do not consider this scenario.}. The last condition is required to ensure  the absence of
GW-induced gradient instabilities \footnote{In the presence of
a sizeable cubic Horndeski operator, dark energy perturbations develop instabilities on gravitational wave backgrounds
as sourced by massive black hole binaries.} \cite{Creminelli:2019kjy}. Thus, we also consider two different sub-types of models living within the propto-Scale class. For the Inv-Hubble-Squared model, instead, we only study the case $\alpha_B\ne0$ with $\alpha_M=0$ \cite{Traykova:2021hbr}. This parametrizaton is inspired by shift-symmetric models \cite{Traykova:2021hbr}. In the cases in which we do not impose the relation $c_M=-c_B$, we force the positivity of the non-zero constant $c_i$ through the corresponding prior to satisfy the stability condition for perturbations.
 
As it will become clear in the next sections, our methodology does not depend on the specific model under consideration, since it can also be applied to more general Horndeski models, or even other models, as far as they introduce modifications from GR at the redshifts covered by the parametrizations described in Sec. \ref{sec:binning_method} and respect the various working assumptions also explained in that section.

\section{Data}\label{sec:data}
In this paper, we consider RSD data expressed in terms of $f(z) \sigma_8(z)$ where $f = d \ln \delta_m/d \ln a$ is the growth rate and $\sigma_8$ is the amplitude of mass fluctuations in spheres of $8 h^{-1}$ Mpc. The quantity $f \sigma_8$ represents the amplitude of the velocity divergence power spectrum that is probed by RSD measurements. 

\subsection{Mock data in the validation analysis and the DESI forecast}\label{sec:dataMock}

In the validation analysis carried out in Sec. \ref{sec:DESI} and the DESI forecast of Sec. \ref{sec:DESIforecast} we employ mock RSD data from DESI. We employ the forecasted relative errors $\frac{\sigma\left(f\sigma_{8}\right)}{f\sigma_{8}}(z_i)$ displayed in Table 7 of \cite{DESI:2023dwi}. The DESI mock data cover the redshift range $z\in (0,2.1)$ in 21 equidistant redshift bins, with central redshifts $z_{i}=0.05+0.1i$ for $i=0,1,...,20$. We perform the validation analysis for the models described in Sec. \ref{sec:EFT} and using different mock data sets. The central values of the mock data are computed using Eq. \eqref{eq:delta_evolve} and setting in all cases $\Omega_m\equiv\Omega_m(z=0)=0.3069$. 
We assume either the $\Lambda$CDM ($c_B=c_M=0$) or the same modified gravity model employed in the validation analysis, with $c_B=0.2,0.4$ if $c_M=0$ or $c_M=-c_B$, or $c_M=0.2,0.4$ if $c_B=0$. To solve Eq. \eqref{eq:delta_evolve}, we assume the initial condition $\delta_m(a_{\rm ini})=\delta_{m,*}\, \left(1+\frac{3a_{\rm ini}}{2a_{\mathrm{eq}}}\right)$ at $a_{\rm ini}=e^{-5}$, setting the matter-radiation equality $a_{\mathrm{eq}}=1/3300$ and  $\delta_{m,*}=2.118\times10^{-4}$. This normalization leads to $\sigma_{8}(z=0)=0.832$ in the $\Lambda$CDM model with  $\Omega_m=0.3069$. The value of $\delta_{\mathrm{m,*}}$ will not affect the parameter constraints in Sec. \ref{sec:DESI} as long as we use (as we do) the same value employed in the generation of the mock data, since it is just an overall factor. 

Since we know the underlying cosmology employed to build the DESI mock data set, we can check if we recover the correct central values of the EFT parameters and $\Omega_m$ from the corresponding fitting analyses of the original and the compressed data. See Sec. \ref{sec:DESI} for details.

\subsection{Current LSS data}

In Table \ref{tab:fs8_table}, we present the list with the 15 RSD data points employed in Sec. \ref{sec:current_data}, together with the corresponding references. They are provided by different galaxy surveys and cover the redshift range $z\in(0.01,1.5)$. This sample is quite conservative, as it avoids double-counting issues. For example, CMASS and WiggleZ use different tracers: CMASS focuses on luminous, primarily red galaxies, whereas WiggleZ targets emission-line galaxies in low-to-intermediate mass halos. Although the two surveys overlap in both sky area and redshift, they remain free from double-counting problems \cite{Marin:2015ula}. CMASS, in fact, covers a much larger area - $10^5$ deg$^2$ compared to WiggleZ’s 816 deg$^2$. Similar considerations apply to the other data points listed in Table \ref{tab:fs8_table}.

\begin{table}[t!]
\begin{center}
\begin{tabular}{| c | c |c | c |}
\multicolumn{1}{c}{Survey} &  \multicolumn{1}{c}{$z$} &  \multicolumn{1}{c}{$f(z)\sigma_8(z)$} & \multicolumn{1}{c}{{\small References}}
\\\hline
ALFALFA & $0.013$ & $0.46\pm 0.06$ & \cite{Avila:2021dqv}
\\\hline
6dFGS+SDSS & $0.035$ & $0.338\pm 0.027$ & \cite{Said:2020epb}
\\\hline
GAMA & $0.18$ & $0.29\pm 0.10$ & \cite{Simpson:2015yfa}
\\ \cline{2-4}& $0.38$ & $0.44\pm0.06$ & \cite{Blake:2013nif}
\\\hline
 WiggleZ & $0.22$ & $0.42\pm 0.07$ & \cite{Blake:2011rj} \tabularnewline
\cline{2-3} & $0.41$ & $0.45\pm0.04$ & \tabularnewline
\cline{2-3} & $0.60$ & $0.43\pm0.04$ & \tabularnewline
\cline{2-3} & $0.78$ & $0.38\pm0.04$ &
\\\hline
DR12 BOSS & $0.32$ & $0.427\pm 0.056$  & \cite{Gil-Marin:2016wya}\\ \cline{2-3}
 & $0.57$ & $0.426\pm 0.029$ &
\\\hline
VIPERS & $0.60$ & $0.49\pm 0.12$ & \cite{Mohammad:2018mdy}
\\ \cline{2-3}& $0.86$ & $0.46\pm0.09$ &
\\\hline
VVDS & $0.77$ & $0.49\pm0.18$ & \cite{Guzzo:2008ac},\cite{Song:2008qt}
\\\hline
FastSound & $1.36$ & $0.482\pm0.116$ & \cite{Okumura:2015lvp}
\\\hline
eBOSS Quasar & $1.48$ & $0.462\pm 0.045$ & \cite{eBOSS:2020gbb}
\\\hline
 \end{tabular}
\end{center}
\caption{Published values of $f(z)\sigma_8(z)$. See the quoted references for further details.}
\label{tab:fs8_table}
\end{table}


\section{Testing the method with mock data from DESI}\label{sec:DESI}

\subsection{Methodology}\label{sec:methodology}
In order to show the robustness of our compression method, we perform these steps for each mock data set and EFT model: 

\begin{enumerate}

\item First, we perform the fitting analysis to constrain the EFT parameters directly from the DESI mock data (see Sec. \ref{sec:dataMock}) using the following $\chi^2$, 

\begin{equation}\label{eq:chi1}
\chi^2(c_j,\Omega_m) = \sum_{i=0}^{20}\left(\frac{f\sigma_{8,{\rm EFT}}(z_i,c_j,\Omega_m)-f\sigma_8(z_i)|_{\rm mock}}{\sigma(z_i)|_{\rm mock}}\right)^2\,.
\end{equation}
In this paper, we will refer to this method as the ``direct method''. We actually study two cases to see what is the impact of $\Omega_m$ on the fitting results. In one of them we fix it to $\Omega_m=0.3069$, and in the other we allow it to vary, using the prior

\begin{equation}\label{eq:priorOm}
    \Omega_m= 0.3069\pm0.0050\quad (68\%\,{\rm C.L.})\,.
\end{equation}

The latter is obtained in $\Lambda$CDM using the DESI BAO data, the CMB temperature and polarization data from {\it Planck} and the CMB lensing data from {\it Planck} and ACT \cite{DESI:2024mwx}. The results of the fitting analyses carried out using Eq. \eqref{eq:chi1} are displayed in Tables \ref{Tab:2}-\ref{Tab:lastfit}. It is important to remark that data on $f\sigma_8$ alone cannot break the degeneracy between $\Omega_m$ and $\mu$, which is apparent from Eq. \eqref{eq:delta_evolve}. This is why we impose a prior on the matter density parameter obtained from  background observations. In this work, we are interested in the effect of varying $\Omega_m$ within the limits of its current background constraints on the constraints on $\mu$. If we avoided the use of a prior on $\Omega_m$, we would be only capable of constraining the product $\Omega_m\mu$ \cite{Zheng:2023yco}.

\item We repeat the same exercise, but now we use the parametrizations of $\mu(z)$ described in Sec. \ref{sec:binning_method} (instead of the EFT model), with 

\begin{equation}
\chi^2(\mu_j,\Omega_m) = \sum_{i=0}^{20}\left(\frac{f\sigma_{8,{\rm bin}}(z_i,\mu_j,\Omega_m)-f\sigma_8(z_i)|_{\rm mock}}{\sigma(z_i)|_{\rm mock}}\right)^2\,.
\end{equation}
This is the ``bin method''. The corresponding constraints are also provided in Tables \ref{Tab:2}-\ref{Tab:lastfit}. 

\item The next step consists in translating the constraints on $(\mu_i,\Omega_m)$ into constraints on $(c_i,\Omega_m)$ and see if they match with those obtained in point 1. For this, we need to obtain the fitting formula of $\mu_i(c_i,\Omega_m)$. This is done by generating curves of $f\sigma_{8,{\rm EFT}}(z_i,c_j,\Omega_m)$ using a grid of values $(c_i,\Omega_m)$ and minimizing 

\begin{equation}\label{eq:dictionary}
\chi^{2}(\mu_{j})=\sum_{i=0}^{20}\left(\frac{f\sigma_{8,\mathrm{EFT}}(z_{i},c_k,\Omega_m)-f\sigma_{8,\mathrm{bin}}(z_{i},\mu_{j},\Omega_m)}{\sigma(z_{i})|_{\rm mock}}\right)^2\,.
\end{equation}
We obtain very low values of the minimum $\chi^2$, $\chi^2_{\rm min}$, from Eq. \eqref{eq:dictionary}. This means that our parametrizations in terms of the $\mu_i$'s are able to reproduce the shapes of $f(z)\sigma_8(z)$ predicted in complex EFTofDE.

This procedure allows us to build a dictionary to relate the values of the EFT parameters with the $\mu_i$'s through the corresponding fitting formulae. We have checked that a quadratic formula of the following type,

\begin{equation}\label{eq:fittingFormula}
\mu_i(c_j,\Omega_m) = 1+c_j[A_i+B_i(\Omega_m-0.3)]+D_ic_j^2    \,,
\end{equation}
with $A_i,B_i,D_i$ fitting parameters, is accurate enough, since it allows us to recover the constraints on EFT parameters at the posterior level. We illustrate this in Fig. \ref{Fig:1}, where it is apparent that the expected uncertainties of the final DESI data are much larger than the error of the fitting formula. See also Sec. \ref{sec:resultsTests}. 

In this expression, $c_j=c_M$ only if $c_B=0$. Otherwise, $c_j=c_B$. In Table \ref{tab:fitting_coefficients}, we provide the values of the fitting parameters for all the EFT models studied in this paper. We employ a $\Lambda$CDM background in most of the cases to derive the fitting formulae, but we also study the impact of the background choice on the shape of formula \eqref{eq:fittingFormula} to make sure that it is indeed small. We discuss these technical and important details in the context of model Inv-Hubble-Squared, see Sec. \ref{sec:resultsInvHubbleSquared}. 

We should note that the initial amplitude of perturbations (i.e. $A_s$ or $\sigma_{8,0}$) is just an overall factor of the solution for the matter density contrast $\delta_m$ and this is why the mapping formula \eqref{eq:fittingFormula} is independent of it. See the comments in Sec. \ref{sec:dataMock}.

\item Finally, to convert the constraints on $\mu_i$ into constraints on $c_{i}$ we
perform a Monte Carlo analysis using the constraints obtained in point 2 as the data and the fitting formula derived in point 3 as the theory input. Assuming a Gaussian distribution, we employ
the $\chi^2$ 
\begin{equation}
\chi^2(c_{j},\Omega_m)=
\left[\boldsymbol{p}(c_{j},\Omega_m)-\boldsymbol{p}_{\mathrm{mean}}\right]^{\mathrm{T}}Cov^{-1}\left[\boldsymbol{p}(c_{j},\Omega_m)-\boldsymbol{p}_{\mathrm{mean}}\right],
\label{eq:likelihood}
\end{equation}
where $\boldsymbol{p}(c_{j},\Omega_m)$ and $\boldsymbol{p}_{\mathrm{mean}}$ are the vectors containing the various $\mu_i$'s and $\Omega_m$ from theory and data, respectively. The covariance matrix $Cov$ is also obtained in point 2. With a prior on $\Omega_m$, we found that the Gaussian approximation for these parameters is valid for the DESI mock data. We have explicitly tested the validity of the Gaussian approximation by using the full posterior of the $\mu_i$'s and comparing the results with those obtained using the Gaussian approximation. We found negligible differences compared to the typical uncertainties. Thus the use of the Gaussian approximation is fully justified, since it simplifies the analysis and does not induce any significant error.

\end{enumerate}

The constraints on $(c_i,\Omega_m)$ obtained in the last step can be directly compared with those obtained in the first step in order to see whether the compressed information leads to unbiased results and preserves the statistical power of the original LSS data. In Sec. \ref{sec:resultsTests} we present our results, considering all the combinations of mock data and models described above.

\begin{table}[t!]
\begin{tabular}{|c|c|c|c|c|c|}
\hline
Model                         & Subtype                             & $\mu_i$ & $A_i$ & $B_i$ & $D_i$ \\ \hline
\multirow{4}{*}{Propto-Omega} & \multirow{2}{*}{$c_B\ne 0$\,;\,$c_M=0$} & $\mu_1$ &   0.0952    &   $-0.419$    &   0.0106    \\ \cline{3-6} 
                              &                                     & $\mu_2$ &   0.0221    &   $-0.105$    &    0.000273   \\ \cline{2-6} 
                              & \multirow{2}{*}{$c_M\ne 0$ \,;\,$c_B=0$} & $\mu_1$ &  0.113     &   $-0.189$    &   $-0.0213$    \\ \cline{3-6} 
                              &                                     & $\mu_2$ &   0.0346    &   $-0.140$    &   $-0.00105$    \\ \hline\hline
\multirow{4}{*}{Inv-Hubble-Squared} & \multirow{2}{*}{$c_B\ne 0$ \,;\,$c_M=0$} & $\mu_1$ &   0.136    &   $-0.396$    &   0.0218    \\ \cline{3-6} 
                              &                                     & $\mu_2$ &   0.0314    & $-0.105$   &    0.00085   \\ \cline{2-6}
&\multirow{1}{*}{$c_B\ne 0$ \,;\,$c_M=0$} & $\mu_1$ &   0.135    &   $-0.390$    &   0.0222    \\ \cline{3-6} 
                              &   \multirow{1}{*}{$w_0$-$w_a$ background}                                  & $\mu_2$ &   0.0319    & $-0.108$   &    0.000733   \\ \hline \hline
\multirow{6}{*}{Propto-Scale} & \multirow{3}{*}{$c_B\ne 0$\,;\,$c_M=0$} & $\mu_1$ &  0.292   &    0.0795  &    0.160   \\ \cline{3-6} 
                              &                                     & $\mu_2$ &   0.289   &   0.0741    &   0.152   \\ \cline{3-6}
                               &                                     & $\mu_3$ &  0.217    &    0.0231   &   0.0128  \\ \cline{2-6} 
                              & \multirow{3}{*}{$c_B=-c_M$} & $\mu_1$ &   0.495   &   $-0.0921$   &   0.0739   \\ \cline{3-6} 
                              &                                     & $\mu_2$ &  0.382   &   $-0.0336$   &   0.0641  \\   \cline{3-6}
                                &                                    & $\mu_3$ &  0.152    &   0.207   &   0.0036   \\ \hline
\end{tabular}
\caption{Values of the fitting parameters entering Eq. \eqref{eq:fittingFormula} obtained for the various EFT models under study. We use a $\Lambda$CDM background unless specified otherwise in the ``Subtype'' column. }
\label{tab:fitting_coefficients}
\end{table}

\subsection{Results}\label{sec:resultsTests}

\subsubsection{propto-Omega}

\begin{figure}[h!]
\includegraphics[height=15.2cm]{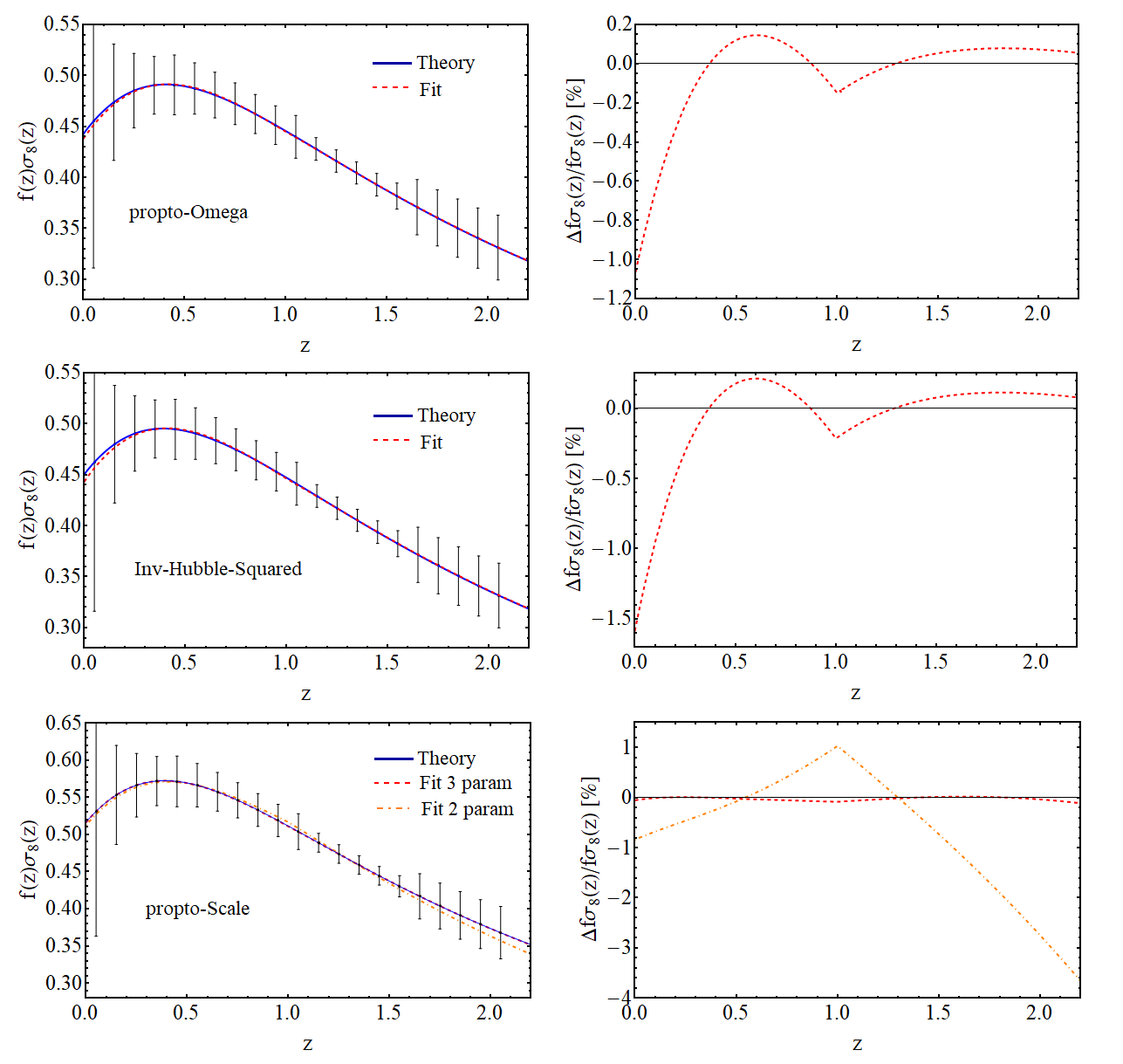}
\caption{Comparison of the theoretical curves of $f(z)\sigma_8(z)$ obtained for the various EFT models employed in this paper with $c_B=0.4$ and $\Omega_m=0.3$ (in blue), and the corresponding best-fit curves of the bin model (in red or orange). The latter have been obtained following the procedure explained in point 3 of Sec. \ref{sec:methodology}. We also show the DESI mock data, as described in Sec. \ref{sec:dataMock}. The differences between the theoretical and fitted curves are negligible compared to the forecasted uncertainties. In the right column we plot the percentage relative difference $\Delta f\sigma_8(z)/f\sigma_8(z)$ [\%] between the theoretical and best-fit results. They are smaller than $\sim 1.5\%$ at $z<0.5$ and smaller than $0.2\%$ at $z>0.5$ in absolute value, with the only exception of the fit of the propto-Scale model using $\mu$-2param. As explained in Sec. \ref{sec:resultsScales}, in this case we need 3 parameters to improve the accuracy of the fitting formula. The sudden change in the derivative of $f(z)\sigma_8(z)$ found at $z=1$, which is apparent in the plots on the right, is expected and caused by the discontinuity of $\mu(z)$ at this redshift, cf. formulas \eqref{eq:2p} and \eqref{eq:3p}}.   
\label{Fig:1}
\end{figure}

\begin{figure}[h!]
\includegraphics[height=15cm]{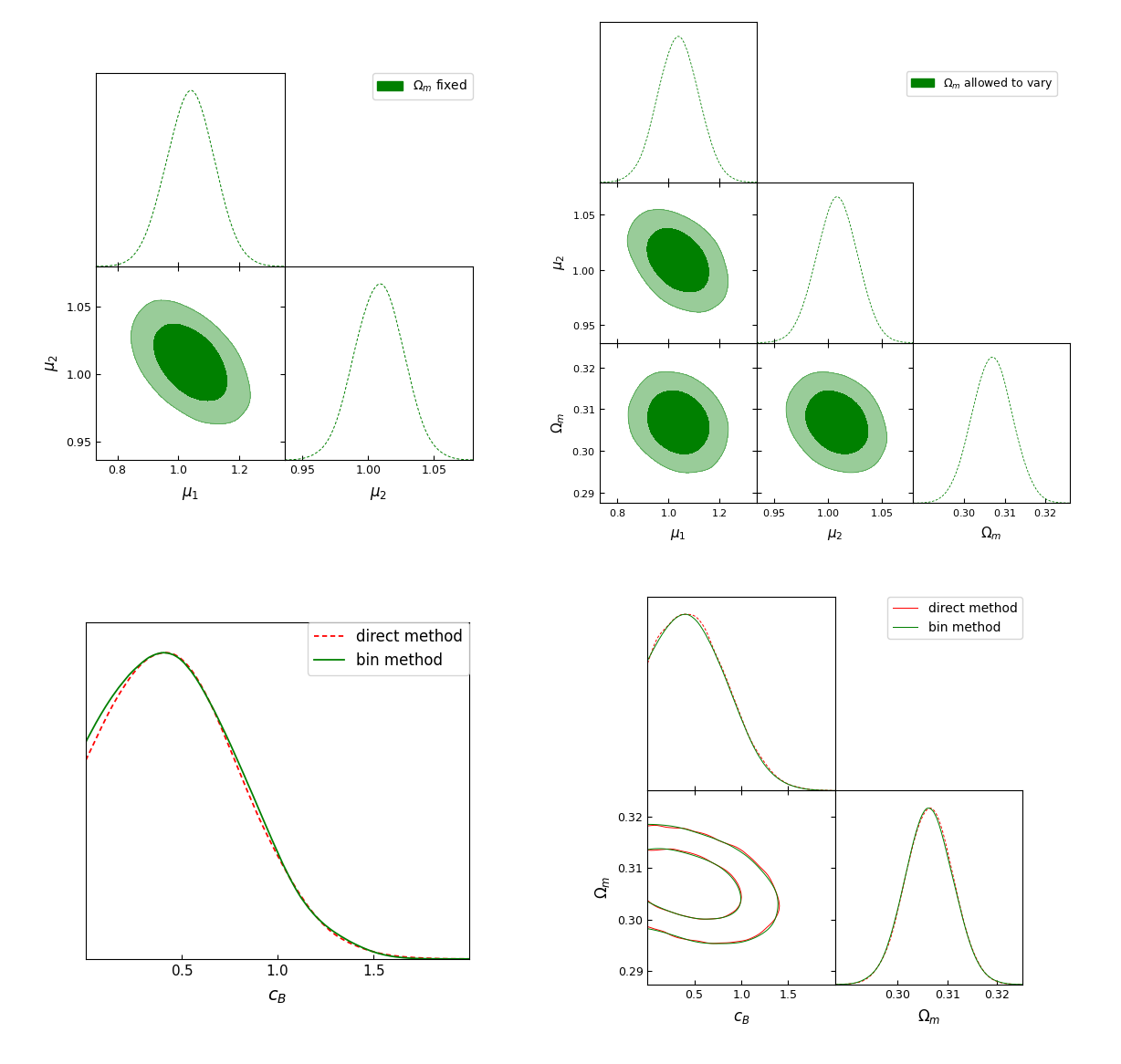}
\caption{{\it Upper plots:} Posterior distributions of $\mu_1$ and $\mu_2$ in the case in which $\Omega_m$ is fixed (left plot) and in which it is allowed to vary in the Monte Carlo (right plot). In that case we also constrain $\Omega_m$, but this constraint is dominated by the prior.  We use the mock data from DESI obtained with $c_B=0.4$ (and $c_M=0$) in the  propto-Omega model; {\it Lower plots:} In the left plot we show the constraints on $c_B$ derived with the direct and bin methods fixing the value of $\Omega_m$. When $\Omega_m$ is allowed to vary we obtain the results of the right plot. The agreement between the results obtained with the direct and bin methods is excellent.}\label{Fig:2}
\end{figure}

In Figure~\ref{Fig:2}, as an example, we illustrate the results more graphically for the case in which we produce the mock data using $c_B=0.4$ (and $c_M=0$) and constrain the model with varying $c_B$. More concretely, we show the contours and one-dimensional posterior distributions obtained for 
the EFT parameters with the direct and bin methods, and also the constraints on the $\mu_i$'s. As in Table \ref{Tab:2}, this figure demonstrates the great consistency between the two results. From the first Monte Carlo analysis of the bin method, i.e., the one carried out to obtain the constraints on the parameters of $\mu$-2param, we cannot only extract the mean values and uncertainties of the latter, but also the covariance matrix, which reads, 

\begin{equation}
Cov=\left(\begin{array}{cc}
\sigma_{\mu_{1}}^{2} & \sigma_{\mu_{1}\mu_{2}}\\
\sigma_{\mu_{1}\mu_{2}} & \sigma_{\mu_{2}}^{2}
\end{array}\right)=\left(\begin{array}{cc}
6.147\times10^{-3} & -6.495\times10^{-4}\\
-6.495\times10^{-4} & 3.413\times10^{-4}
\end{array}\right).
\label{eq:cov}
\end{equation}
This is the covariance matrix that enters Eq. \eqref{eq:likelihood} and is employed in the second Monte Carlo run to obtain the projected constraints on the EFT parameters in the bin method. Figure \ref{Fig:2} demonstrates that the Gaussian approximation is sufficient, since the two-dimensional contours are very close to perfect ellipses. While we present the mean values and errors of the $\mu_i$'s from all analyses in this work in the various tables, we provide the covariance matrix only for this particular case to maintain compactness.

Table \ref{Tab:2} shows an excellent consistency between the two sets of constraints regardless of the mock data and model subtype employed in the analysis. We can conclude that the compression of the statistical content of the RSD data in terms of the parameters $\mu_i$ is carried out very efficiently. In addition, these results also demonstrate the good performance of the fitting function provided in Eq. \eqref{eq:fittingFormula}.

\begin{table}[t!]
\[
\begin{array}{cc|ccc|cccc}
\hline 
c_{B} &  & 0.0 & 0.2 & 0.4 &  & 0.0 & 0.2 & 0.4\\
\hline\hline \mathrm{Direct} & c_{B} & <0.8175 & <0.9366 & <1.0900 &  & <0.8696 & <0.9858 & <1.1399\\
 & c_{B} & 0.283_{-0.199}^{+0.308} & 0.366_{-0.246}^{+0.331} & 0.484_{-0.301}^{+0.364} &  & 0.309_{-0.214}^{+0.322} & 0.389_{-0.263}^{+0.349} & 0.500_{-0.316}^{+0.381}\\
 & \Omega_{m} & - & - & - &  & 0.306\pm0.005 & 0.306\pm0.005 & 0.306\pm0.005\\
\hline\hline \mathrm{Bin} & \mu_{1} & 1.001\pm0.077 & 1.019_{-0.077}^{+0.078} & 1.039_{-0.079}^{+0.076} &  & 1.002_{-0.079}^{+0.078} & 1.019\pm0.079 & 1.039_{-0.080}^{+0.081}\\
 & \mu_{2} & 1.000\pm0.018 & 1.004\pm0.018 & 1.008_{-0.018}^{+0.018} &  & 0.994\pm0.019 & 1.004_{-0.018}^{+0.019} & 1.008\pm0.019\\
 & \Omega_{m} & - & - & - &  & 0.307\pm0.005 & 0.307\pm0.005 & 0.307\pm0.005\\
\hline  & c_{B} & <0.8156 & <0.9571 & <1.1124 &  & <0.8657 & <0.9930 & <1.1377\\
 & c_{B} & 0.289_{-0.201}^{+0.302} & 0.378_{-0.254}^{+0.338} & 0.493_{-0.310}^{+0.374} &  & 0.305_{-0.212}^{+0.322} & 0.387_{-0.260}^{+0.356} & 0.491_{-0.316}^{+0.386}\\
 & \Omega_{m} & - & - & - &  & 0.305\pm0.005 & 0.306\pm0.005 & 0.306\pm0.005\\
\hline
\end{array}
\]
\[
\begin{array}{cc|ccc|cccc}
\hline 
c_{M} &  & 0.0 & 0.2 & 0.4 &  & 0.0 & 0.2 & 0.4\\
\hline\hline \mathrm{Direct} & c_{M} & <0.6866 & <0.8675 & <1.0586 &  & <0.7538 & <0.9157 & <1.1154\\
 & c_{M} & 0.227_{-0.160}^{+0.261} & 0.334_{-0.224}^{+0.307} & 0.471_{-0.282}^{+0.344} &  & 0.254_{-0.178}^{+0.284} & 0.352_{-0.236}^{+0.320} & 0.484_{-0.296}^{+0.370}\\
 & \Omega_{m} & - & - & - &  & 0.305\pm0.005 & 0.306\pm0.005 & 0.306\pm0.005\\
\hline\hline \mathrm{Bin} & \mu_{1} & \mathrm{same} & 1.021\pm0.078 & 1.042_{-0.079}^{+0.080} &  & \mathrm{same} & 1.022\pm0.080 & 1.040_{-0.078}^{+0.079}\\
 & \mu_{2} & \mathrm{as}\,c_M=0 & 1.006\pm0.018 & 1.013_{-0.019}^{+0.018} &  & \mathrm{as}\, c_M=0 & 1.007\pm0.019 & 1.013\pm0.019\\
 & \Omega_{m} & - & - & - &  &  & 0.307\pm0.005 & 0.307\pm0.005\\
\hline  & c_{M} & <0.6916 & <0.8579 & <1.0653 &  & <0.7521 & <0.9256 & <1.1310\\
 & c_{M} & 0.227_{-0.160}^{+0.259} & 0.330_{-0.218}^{+0.305} & 0.470_{-0.280}^{+0.345} &  & 0.254_{-0.176}^{+0.279} & 0.354_{-0.236}^{+0.327} & 0.489_{-0.300}^{+0.371}\\
 & \Omega_{m} & - & - & - &  & 0.305\pm0.005 & 0.306\pm0.005 & 0.306\pm0.005\\
\hline
\end{array}
\]
\caption{ MCMC results obtained with the direct and bin methods for the propto-Omega models studied in this paper. In the upper block we show the results obtained considering $c_B\ne 0$ (with $c_M=0$) and in the lower block those obtained considering $c_M\ne 0$ (with $c_B=0$). For the $c_i$'s we display the results at 68\% and 95\% confidence levels in the second and first rows, respectively. We fix $\Omega_{m}=0.3069$ in the left side of
the table, while we also sample $\Omega_{m}$ in the right side of the table. We use a $\Lambda$CDM background.  
\label{Tab:2} 
} \end{table}

\subsubsection{Inv-Hubble-Squared}\label{sec:resultsInvHubbleSquared}
We show the results on Inv-Hubble-Squared model for varying $c_B$ (and $c_M=0$) in Table~\ref{Tab:4} and Figure~\ref{Fig:1}. We also employ in this case two bins for $\mu$ in the bin method. Again, the results obtained with the two approaches are in excellent agreement.  The  parametrisation of the Inv-Hubble-Squared model was constructed based on the scaling solution that modifies the background expansion history from $\Lambda$CDM \cite{Traykova:2021hbr}.
Thus we test whether the fitting formula is sensitive to the background or not in this case. 
We note that we include the effect of dynamical dark energy only in the background in our analysis. 
Assuming the mean value of $w_0 w_a$DE model from CMB+DESI+SNIa(Panth.) as a background theory ($w_{0}=-0.827$
and $w_{a}=-0.75$), we found that the impact of the different background on the fitting formula Eq. \eqref{eq:fittingFormula} is negligible, as it is clear from the numbers displayed in Table \ref{tab:fitting_coefficients}. To quantify the change in our results induced by the modification of the background, we create the mock data using this background and repeat the analysis as shown in Table \ref{Tab:7}. The constraints are almost unchanged from the case of the $\Lambda$CDM background. We then assess the bias of using the fitting formula calibrated in the $\Lambda$CDM background when we convert the constraints on $\mu_i$ to $c_B$. The result is also shown in Table \ref{Tab:7}. As we can see the constraints are almost unaffected by the use of the fitting formula obtained in the $\Lambda$CDM background.

\subsubsection{propto-Scale}\label{sec:resultsScales}

In the case the $\alpha_i$ parameter is proportional to the scale factor, the effect of modified gravity appears at higher redshifts. This results in inaccuracies of the 2 bins fit where we obtained $\Delta \chi^2_{\mathrm{2Bin}}=0.615$ for $c_B=0.4$, with $\chi^2$ defined in Eq. (\ref{eq:dictionary}). In this case, the third bin is required. Using 3 bins, we obtain $\Delta \chi^2_{\mathrm{3Bin}}=0.002$. The significantly improved agreement between the theoretical curve and the fitting result is evident in Figure~\ref{Fig:1}.

We show the results on propto-Scale model for varying $c_B$ in Table~\ref{Tab:lastfit}. Since the modification of gravity appears early in this model, the constraint on $c_B$ is much tighter than the other two models.

In order to make a quantitative comparison between the bin and direct methods 
we employ the following expression,
\begin{equation}\label{eq:tensionT}
T\equiv\frac{\left|c_{j,\mathrm{Direct}}-c_{j,\mathrm{\mu}}\right|}{\sqrt{\sigma_{c_{j,\mathrm{Direct}}}^{2}+\sigma_{c_{j,\mathrm{\mu}}}^{2}}}\,,
\end{equation}
with $j=B$ (if $c_M=0$) or $j=M$ (if $c_B=0$). We calculate the tension $T$ using  our constraints at 68\% C.L. 
We find that the tension $T<0.17$ in all cases, which means that the bias of the 1D marginalised constraint on the model parameter due to the use of the binning is at most 0.17$\sigma$ for the final year DESI data. This systematic error can be estimated from the synthetic data analysis for any given model, and this can be included in the error budget. We also note that the current data prefers a lower value of $\mu_1$ than the prediction of $\Lambda$CDM with significance greater than 2$\sigma$ as shown in the following section and the significance is expected to be greater in DESI if the best-fit value remains the same.

\begin{table}[h!]
\[
\begin{array}{cc|ccc|cccc}
\hline 
c_{B} &  & 0.0 & 0.2 & 0.4 &  & 0.0 & 0.2 & 0.4\\
\hline\hline \mathrm{Direct} & c_{B} & <0.5597 & <0.6925 & <0.8575 &  & <0.6078 & <0.7381 & <0.8904\\
 & c_{B} & 0.198_{-0.138}^{+0.209} & 0.286_{-0.185}^{+0.243} & 0.429_{-0.246}^{+0.263} &  & 0.215_{-0.153}^{+0.227} & 0.299_{-0.196}^{+0.258} & 0.431_{-0.255}^{+0.280}\\
 & \Omega_{m} & - & - & - &  & 0.305\pm0.005 & 0.306\pm0.005 & 0.307\pm0.005\\
\hline\hline \mathrm{Bin} & \mu_{1} & \mathrm{same} & 1.027\pm0.078 & 1.058_{-0.080}^{+0.077} &  & \mathrm{same} & 1.028_{-0.078}^{+0.080} & 1.059_{-0.080}^{+0.081}\\
 & \mu_{2} & \mathrm{as\,}c_{M}=0 & 1.006\pm0.018 & 1.012_{-0.019}^{+0.018} &  & \mathrm{as\,}c_{M}=0 & 1.006\pm0.019 & 1.012\pm0.019\\
 & \Omega_{m} & \textrm{on\,propto-Omega} & - & - &  & \textrm{on\,propto-Omega} & 0.307\pm0.005 & 0.307\pm0.005\\
\hline  & c_{B} & <0.5619 & <0.6967 & <0.8578 &  & <0.6167 & <0.7468 & <0.8924\\
 & c_{B} & 0.197_{-0.138}^{+0.209} & 0.285_{-0.188}^{+0.244} & 0.419_{-0.240}^{+0.266} &  & 0.219_{-0.152}^{+0.230} & 0.309_{-0.204}^{+0.260} & 0.432_{-0.253}^{+0.280}\\
 & \Omega_{m} & - & - & - &  & 0.305\pm0.005 & 0.306\pm0.005 & 0.307\pm0.005\\
\hline
\end{array}
\]
\caption{Same as Table~\ref{Tab:2}, but for the Inv-Hubble-Squared model with $c_B\ne 0$ and $c_M=0$. We assume the $\Lambda$CDM model as a background. \label{Tab:4}}
\end{table}

\begin{table}[h!]
\[
\begin{array}{cc|ccc|cccc}
\hline 
c_{B} &  & 0.0 & 0.2 & 0.4 &  & 0.0 & 0.2 & 0.4\\
\hline\hline \mathrm{Direct} & c_{B} & <0.5609 & <0.6937 & <0.8563 &  & <0.5910 & <0.7265 & <0.8785\\
 & c_{B} & 0.198_{-0.138}^{+0.208} & 0.290_{-0.190}^{+0.244} & 0.418_{-0.267}^{+0.239} &  & 0.209_{-0.146}^{+0.217} & 0.298_{-0.196}^{+0.252} & 0.431_{-0.250}^{+0.271}\\
 & \Omega_{m} & - & - & - &  & 0.306\pm0.005 & 0.306\pm0.005 & 0.307\pm0.005\\
\hline\hline \mathrm{Bin} & \mu_{1} & 0.999^{+0.081}_{-0.079} & 1.028_{-0.080}^{+0.079} & 1.057\pm0.080 &  & 1.001^{+0.080}_{-0.079} & 1.028\pm0.082 & 1.057\pm0.082\\
 & \mu_{2} & 1.000\pm0.018 & 1.006_{-0.019}^{+0.018} & 1.012_{-0.019}^{+0.018} &  & 1.000^{+0.018}_{-0.019} & 1.006\pm0.019 & 1.013_{-0.019}^{+0.018}\\
 & \Omega_{m} & - & - & - &  & 0.307\pm0.005 & 0.307\pm0.005 & 0.307\pm0.005\\
\hline  
\multicolumn{9}{c}{\textrm{Fitting formula in the $w_0 w_a$ DE background}} \\ 
\hline
& c_{B} & <0.5631 & <0.7096 & <0.8617 &  & <0.6014 & <0.7452 & <0.8837\\
 & c_{B} & 0.197_{-0.139}^{+0.210} & 0.291_{-0.190}^{+0.247} & 0.423_{-0.244}^{+0.266} &  & 0.211_{-0.147}^{+0.224} & 0.305_{-0.203}^{+0.259} & 0.428_{-0.252}^{+0.277}\\
 & \Omega_{m} & - & - & - &  & 0.306\pm0.005 & 0.307\pm0.005 & 0.307\pm0.005\\
\hline
\multicolumn{9}{c}{\textrm{Fitting formula in the $\Lambda$CDM background}} \\ 
\hline  & c_{B} & <0.5692 & <0.7029 & <0.8694 &  & <0.6076 & <0.7281 & <0.8887\\
 & c_{B} & 0.198_{-0.139}^{+0.211} & 0.288_{-0.189}^{+0.245} & 0.425_{-0.245}^{+0.269} &  & 0.212_{-0.148}^{+0.225} & 0.296_{-0.193}^{+0.254} & 0.424_{-0.251}^{+0.280}\\
 & \Omega_{m} & - & - & - &  & 0.306\pm0.005 & 0.306\pm0.005 & 0.307\pm0.005\\
\hline
\end{array}
\]
\caption{Same as Table~\ref{Tab:4}, but assuming the $w_0w_a$DE model as a background. See Sec. \ref{sec:resultsInvHubbleSquared} for details. We show the constraints on $c_B$ derived from the constraints from $\mu$ using the fitting formula in the 
$w_0w_a$DE and $\Lambda$CDM backgrounds. 
}\label{Tab:7} 
\end{table}

\begin{table}[h!]
\[
\begin{array}{cc|ccc|cccc}
\hline 
c_{B} &  & 0.0 & 0.2 & 0.4 &  & 0.0 & 0.2 & 0.4\\
\hline\hline \mathrm{Direct} & c_{B} & <0.0546 & <0.2430 & <0.4403 &  & <0.0591 & <0.2471 & <0.4452\\
 & c_{B} & 0.019_{-0.013}^{+0.020} & 0.200_{-0.027}^{+0.026} & 0.399_{-0.026}^{+0.025} &  & 0.020_{-0.014}^{+0.022} & 0.199_{-0.029}^{+0.029} & 0.399_{-0.028}^{+0.028}\\
 & \Omega_{m} & - & - & - &  & 0.305\pm0.005 & 0.307\pm0.005 & 0.307\pm0.005\\
\hline\hline \mathrm{Bin} & \mu_{1} & 0.999_{-0.079}^{+0.078} & 1.067_{-0.081}^{+0.082} & 1.145_{-0.084}^{+0.083} &  & 1.000_{-0.079}^{+0.081} & 1.068_{-0.082}^{+0.082} & 1.144_{-0.085}^{+0.086}\\
 & \mu_{2} & 1.007_{-0.107}^{+0.120} & 1.078_{-0.116}^{+0.123} & 1.152_{-0.122}^{+0.131} &  & 1.010_{-0.109}^{+0.118} & 1.077_{-0.112}^{+0.121} & 1.154_{-0.118}^{+0.130}\\
 & \mu_{3} & 0.994_{-0.107}^{+0.100} & 1.033_{-0.107}^{+0.106} & 1.078_{-0.113}^{+0.111} &  & 0.991_{-0.104}^{+0.103} & 1.033_{-0.106}^{+0.103} & 1.077_{-0.112}^{+0.108}\\
 & \Omega_{m} & - & - & - &  & 0.307\pm0.005 & 0.307\pm0.005 & 0.307\pm0.005\\
\hline  & c_{B} & <0.0612 & <0.2531 & <0.4541 &  & <0.0651 & <0.2563 & <0.4586\\
 & c_{B} & 0.022_{-0.015}^{+0.023} & 0.206_{-0.029}^{+0.028} & 0.403_{-0.032}^{+0.031} &  & 0.024_{-0.016}^{+0.024} & 0.206\pm0.031 & 0.403_{-0.034}^{+0.034}\\
 & \Omega_{m} & - & - & - &  & 0.306\pm0.005 & 0.307\pm0.005 & 0.307\pm0.005\\
\hline\hline \mathrm{Tension} & T & 0.117 & 0.154 & 0.099 &  & 0.145 & 0.165& 0.091\\
\hline
\end{array}
\]

\[
\begin{array}{cc|ccc|cccc}
\hline 
c_{M}=-c_{B} &  & 0.0 & 0.2 & 0.4 &  & 0.0 & 0.2 & 0.4\\
\hline\hline \mathrm{Direct} & c_{M} & <0.0531 & <0.2467 & <0.4464 &  & <0.0572 & <0.2489 & <0.4488\\
 & c_{M} & 0.018^{+0.019}_{-0.013} & 0.201^{+0.028}_{-0.027} & 0.400\pm0.028 &  & 0.020^{+0.022}_{-0.014} & 0.200\pm0.029 & 0.400\pm0.029\\
 & \Omega_{m} & - & - & - &  & 0.308\pm0.005 & 0.307\pm0.005 & 0.307\pm0.005\\
\hline\hline \mathrm{Bin} & \mu_{1} &  & 0.898_{-0.075}^{+0.073} & 0.807\pm0.069 &  &  & 0.898_{-0.076}^{+0.075} & 0.807_{-0.073}^{+0.070}\\
 & \mu_{2} & \mathrm{same} & 0.934_{-0.101}^{+0.109} & 0.865_{-0.098}^{+0.104} &  & \mathrm{same} & 0.933_{-0.101}^{+0.108} & 0.863_{-0.098}^{+0.102}\\
 & \mu_{3} & \mathrm{as} \, c_M=0& 0.962_{-0.099}^{+0.095} & 0.932\pm0.096 &  & \mathrm{as} \, c_M=0 & 0.963_{-0.098}^{+0.096} & 0.934\pm0.095\\
 & \Omega_{m} &  & - & - &  &  & 0.307\pm0.005 & 0.307\pm0.005\\
\hline  & c_{M} & <0.0524 & <0.2471 & <0.4592 &  & <0.0561 & <0.2509 & <0.4644\\
 & c_{M} & 0.017^{+0.020}_{-0.012} & 0.197\pm0.030 & 0.395^{+0.039}_{-0.038} &  & 0.019^{+0.021}_{-0.013} & 0.196\pm0.033 & 0.394\pm0.042\\
 & \Omega_{m} & - & - & - &  & 0.308\pm0.005 & 0.307\pm0.005 & 0.307\pm0.005\\
\hline\hline \mathrm{Tension} & T & 0.043 & 0.098 & 0.105 &  & 0.0394 & 0.091 & 0.118 \\
\hline
\end{array}
\]
\caption{Same as Table \ref{Tab:2}, but for the two propto-Scale models studied in this paper. On the top we show the results obtained considering $c_B\ne 0$ and $c_M=0$, while on the bottom we display the results obtained when $c_{B}=-c_M$. We use a $\Lambda$CDM background. In the last row, we quantify the discrepancy between the values of $c_B$ obtained from the direct and binning methods making use of Eq. \eqref{eq:tensionT}. In all cases $T<0.17$, which demonstrates the good performance of our approach.}
\label{Tab:lastfit}
\end{table}


\section{Applying the formalism to current data and forecast for DESI}\label{sec:VI}

\subsection{Current data}\label{sec:current_data}
In this section we first  constrain the parameters $\mu_i$ of $\mu$-2param and $\mu$-3param using current data on the observable $f(z_i)\sigma_8(z_i)$. We display our LSS data set in Table \ref{tab:fs8_table}. 

In the derivation of the fitting formulae $\mu_i(\Omega_m,c_B)$ and $\mu_i(\Omega_m,c_M)$ displayed in Sec. \ref{sec:DESI} we used Eq. \eqref{eq:delta_evolve} considering a fixed primordial power spectrum. This is legitimate because the $\mu_i$'s control the growth of perturbations in the late-time universe. Moreover, in our validation tests we were not interested in propagating the uncertainties of $A_s$ and $n_s$, since we only aimed to verify the correct performance of the method, and this can be done with fixed values of these parameters. However, if we want to extract meaningful constraints on the parameters $\mu_i$ we need to account for the propagation of the uncertainties of the parameters entering the primordial power spectrum. We use in this part of the analysis a version of the Einstein-Boltzmann code \texttt{CLASS} \cite{Lesgourgues:2011re,Blas:2011rf} that implements the modified linear perturbation equations, incorporating the effect of $\mu\ne 1$. See Appendix \ref{sec:AppendixA} for details.

We construct the binned shapes of $\mu(z)$ considering continuous functions built with hyperbolic tangents. The continuity property is crucial to obtain the correct solution of the matter perturbations from the Einstein-Boltzmann system of equations. This is clear from Eq. \eqref{eq:dotphi}, since it also depends on the derivative of $\mu(z)$. The parametrization $\mu$-2param can be implemented using the following continuous function, 

\begin{equation}\label{eq:muI}
\mu_{\rm 2p}(a) = 1+\frac{1}{2}\left[ \mu_2+\left(\frac{\mu_1-\mu_2}{2}\right)\left(1+\tanh\{\beta(2-1/a)\}\right)-1\right] \times \left[1+\tanh\{\beta(4-1/a)\}\right]\,.
\end{equation}
The parameter $\beta$ appearing in Eq. \eqref{eq:muI} controls the speed of the transitions between the various bins and it can be safely fixed, e.g., to $\beta=5$. For the parametrization $\mu$-3param, instead, we use 

\begin{figure}[t!]
    \centering
    \includegraphics[scale=0.8]{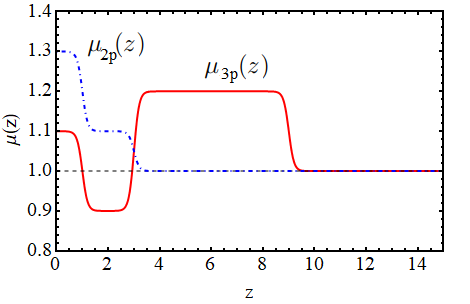}
    \caption{Example of $\mu_{\rm 2p}(z)$ with $\mu_1=1.3$ and $\mu_2=1.1$, and of $\mu_{\rm 3p}(z)$ with $\mu_1=1.1$, $\mu_2=0.9$ and $\mu_3=1.2$. In both cases we have set $\beta=5$. See Eqs. \eqref{eq:muI} and \eqref{eq:muII}.}
    \label{fig:mu(z)}
\end{figure}%

\begin{equation}\label{eq:muII}
\mu_{\rm 3p}(a) =  1+\frac{1}{2}(\tilde{\mu}(a)-1)\left[1+\tanh\{\beta(10-1/a)\}\right]\,,
\end{equation}
with 
\begin{equation}
\tilde{\mu}(a) = \mu_3+\frac{1}{2}\left[ \mu_2+\left(\frac{\mu_1-\mu_2}{2}\right)\left(1+\tanh\{\beta(2-1/a)\}\right)-\mu_3\right] \times \left[1+\tanh\{\beta(4-1/a)\}\right]\,.
\end{equation}
A couple of examples of $\mu_{\rm 2p}(z)$ and $\mu_{\rm 3p}(z)$ are provided in Fig. \ref{fig:mu(z)}. 

As already mentioned, in order to obtain sensible constraints on the parameters $\mu_i$ we need to account for the effect of other parameters that are also relevant for the structure formation processes - such as $\Omega_m$, $A_s$ and $n_s$. Thus, we also vary them in the Monte Carlo analysis and impose some priors on these parameters to keep them in a realistic region of the parameter space and break some existing degeneracies. For $\Omega_m$ we use the Gaussian prior of Eq. \eqref{eq:priorOm}. If we avoided the prior on $\Omega_m$, we could only constrain the product $\mu_i\Omega_m$ \cite{Zheng:2023yco}. On the other hand, we also employ a multivariate Gaussian prior on $\ln(10^{10}A_s)$ and $n_s$ characterized by the following mean vector and covariance matrix,

\begin{equation}
    (\ln(10^{10}A_s),n_s)= (3.044, 0.965)\qquad ;\qquad C[\ln(10^{10}A_s),n_s]=\begin{pmatrix}
2.026\cdot 10^{-4} & 1.384\cdot 10^{-5} \\
1.384\cdot 10^{-5} & 1.735\cdot10^{-5} 
\end{pmatrix}\,.
\end{equation}

\begin{table}[t!]
\begin{center}
\begin{tabular}{| c | c |c | c|}
\multicolumn{1}{c}{Parameters} &  \multicolumn{1}{c}{$\mu$-2param} &  \multicolumn{1}{c}{$\mu$-3param}  &  \multicolumn{1}{c}{$\Lambda$CDM } 
\\\hline 
$\mu_1$ & $0.63\pm 0.14$ (0.66) &  $0.62^{+0.13}_{-0.14}$ (0.58)  & -
\\\hline
$\mu_2$ & $1.10\pm 0.09$ (1.09) &  $0.70^{+0.20}_{-0.42}$ (0.28)  & -
\\\hline
$\mu_3$ & - & $>1.35$ (1.99)  & -\\\hline
$\Omega_m$ & $0.307\pm 0.005$ (0.306) &  $0.307\pm 0.005$ (0.306) & $0.301\pm 0.005$ (0.301) 
\\\hline
$\ln(10^{10}A_s)$ & $3.044^{+0.014}_{-0.015}$ (3.045)  & $3.044\pm 0.015$ (3.046)  & $3.038\pm 0.014$ (3.038)
\\\hline
$n_s$ & $0.965\pm 0.004$ (0.965) &  $0.965\pm 0.004$ (0.966)  & 
$0.964\pm 0.004$ (0.964)
\\\hline
$\chi_{\rm min}^2$ & 12.01 & 8.04  & 20.40
\\\hline
 \end{tabular}
\end{center}
\caption{Mean values and corresponding uncertainties at $68\%$ C.L., together with the best-fit values in brackets, obtained in the analysis with current data. In the last row we show the minimum $\chi^2$, i.e., $\chi^2_{\rm min}$. The constraints on $\Omega_m$, $\ln(10^{10}A_s)$ and $n_s$ are dominated by the priors. See Sec. \ref{sec:current_data} for details.}
\label{tab:fitmu}
\end{table}

\noindent We have extracted this prior from the {\it Planck} analysis of $\Lambda$CDM \cite{Planck:2018vyg}. More concretely, this information can be directly obtained from the Markov chains that are available in the {\it Planck} legacy archive \footnote{https://pla.esac.esa.int/pla/\#cosmology}. This prior is expected to work well for models that do not introduce new physics at very high redshifts, like those considered in this paper. In the Monte Carlo analyses we also employ the flat priors $\mu_i\in [0,2]$ to not explore unphysical regions of the parameter space with negative values of the gravitational coupling or an exaggerated growth of matter perturbations. 

\begin{figure}[h!]
    \centering
    \includegraphics[scale=0.7]{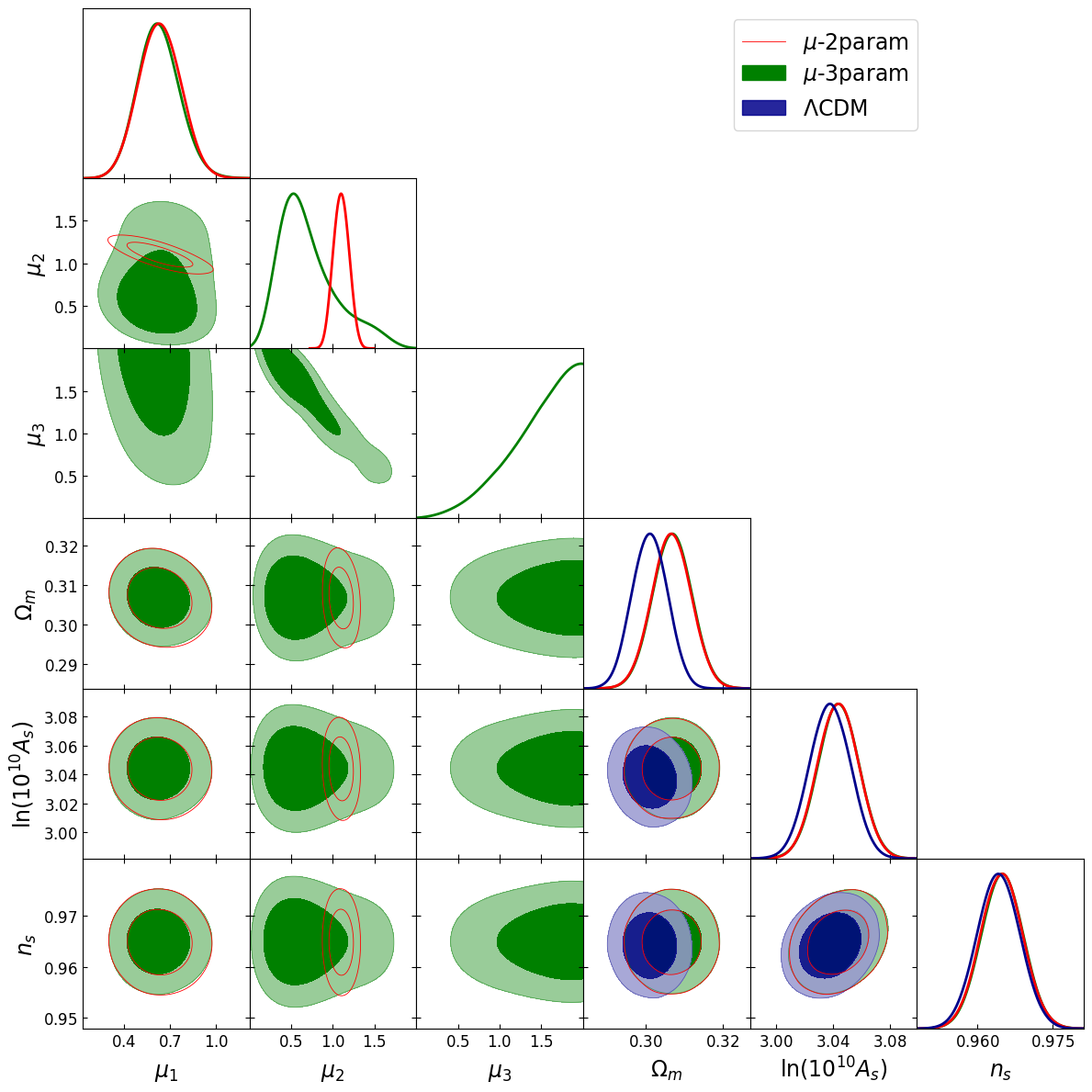}
    \caption{Contour plots and one-dimensional posterior distributions at 68\% and 95\% C.L. obtained from the fitting analyses of $\mu$-2param, $\mu$-3param and the $\Lambda$CDM with current data. }
    \label{fig:triangle_plot}
\end{figure}

\begin{figure}[h!]
   \centering
    \includegraphics[scale=0.6]{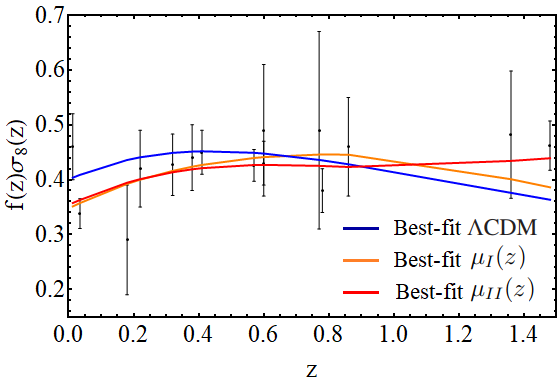}
    \caption{Best-fit curves of $f(z)\sigma_8(z)$ obtained using the two parametrizations $\mu_{\rm 2p}(z)$ and $\mu_{\rm 3p}(z)$, Eqs. \eqref{eq:muI} and \eqref{eq:muII}, together with the data points of Table \ref{tab:fs8_table} and the curve obtained using the $\Lambda$CDM best-fit model.}
    \label{fig:fsig8_bestfit}
\end{figure}

Our results are provided in Table \ref{tab:fitmu} and Fig. \ref{fig:triangle_plot}. In Fig. \ref{fig:fsig8_bestfit} we compare the best-fit curves of $f(z)\sigma_8(z)$ obtained with $\mu$-2param, $\mu$-3param and the $\Lambda$CDM. We find $\mu_1<1$ at $\sim 2.6-2.8\sigma$ C.L. with both parametrizations. The data on $f(z_i)\sigma_8(z_i)$ prefer a lower amount of structure formation in the universe than in $\Lambda$CDM at $z<1$. This explains why our parametrizations are capable of alleviating the tension and decreasing significantly the values of $\chi^2_{\rm min}$. Notice that 13 out of the 15 data points on $f(z_i)\sigma_8(z_i)$ employed in the analysis are inside that redshift range. These results are aligned with those reported in previous works in the literature, see e.g. \cite{Macaulay:2013swa,Joudaki:2017zdt,Gomez-Valent:2017idt,Nesseris:2017vor,Gomez-Valent:2018nib,Benisty:2020kdt,Wright:2020ppw,Nunes:2021ipq,Nguyen:2023fip,Adil:2023jtu} and references therein. At redshifts $z>1$ we only have two data points, with larger central values than in $\Lambda$CDM. This leads to values of $\mu_2$ in the parametrization $\mu$-2param slightly greater than $1$, $\mu_2=1.10\pm 0.09$, but compatible with the standard value ($\mu=1$) at $\sim 1\sigma$ C.L. In $\mu$-3param we find a significant degeneracy in the plane $\mu_2-\mu_3$ (cf. Fig. \ref{fig:triangle_plot}) due to the lack of data points at $z>1$. Current data are not able to set strong constraints on these two parameters simultaneously, and actually the posterior distributions hit the prior boundaries. 

The constraints for  $\mu_1$, $\mu_2$ and $\Omega_m$ obtained with $\mu$-2param are Gaussian in very good approximation \footnote{If deviations from Gaussianity were important, one could for instance reconstruct the Likelihood using the method presented in  \cite{Amendola:2020qkb}, or use a grid-based approach.}. Their central values and uncertainties are provided in Table \ref{tab:fitmu}, and the corresponding covariance matrix reads, 

\begin{equation}\label{eq:cov_current_data} 
C[\mu_1, \mu_2,\Omega_m]=\begin{pmatrix}
1.85\cdot 10^{-2} & -9.64\cdot 10^{-3} & -9.84\cdot 10^{-5}\\
-9.64\cdot 10^{-3} & 8.62\cdot 10^{-3} & -7.85\cdot 10^{-5}\\
-9.84\cdot 10^{-5} &  -7.85\cdot 10^{-5}& 2.46\cdot 10^{-5}
\end{pmatrix}\,.
\end{equation}
Due to the aforementioned strong degeneracy in the $\mu_2-\mu_3$ plane found in $\mu$-3param, we avoid the use of these constraints in the current work.

In this analysis, we have considered two massless neutrinos and one massive neutrino with $m_\nu=0.06$ eV. We have checked that the results remain very stable under reasonable changes of $m_\nu$. For instance, setting $m_\nu=0.10$ eV yields $\mu_1=0.66\pm 0.15$ and $\mu_2=1.10\pm 0.11$; with $m_\nu=0.15$ eV, we obtain $\mu_1=0.67^{+0.15}_{-0.16}$ and $\mu_2=1.13\pm 0.11$. Thus, both the central values and the uncertainties do not show significant shifts.

In addition, we have also tested the stability of the results when we allow $H_0$ to vary in the Monte Carlo analysis. We find a very high anticorrelation between $H_0$ and $\mu_2$, but if we impose a reasonable flat prior $H_0\in[65,75]$ km/s/Mpc the constraints remain again close to those obtained in the main analysis, specially for $\mu_1$: $\mu_1=0.64^{+0.14}_{-0.15}$ and $\mu_2=0.95^{+0.13}_{-0.17}$. The value of $\mu_2$, though, decreases by $\sim 0.8\sigma$ and its uncertainty grows by $\sim 60\%$.

We have projected our constraints onto the parameters of the propto-Omega and Inv-Hubble-Squared models in the cases in which we set $c_M=0$, making use of the Likelihood \eqref{eq:likelihood} and the mapping formulas derived in Sec. \ref{sec:methodology}. For the propto-Omega model we obtain,
\begin{align}\label{eq:con1}
    c_B=0.155^{+0.222}_{-0.114}\,(\mathrm{68\%}\,{\rm C.L.})&\quad ;\quad c_B<0.579\,(\mathrm{95\%}\,{\rm C.L.})\,.     
\end{align}
For the Inv-Hubble-Squared model we are led to  
\begin{equation}
c_B=0.108^{+0.165}_{-0.080}\,(\mathrm{68\%}\,{\rm C.L.})\quad ;\quad c_B
    <0.413\,(\mathrm{95\%}\,{\rm C.L.})\,.
\end{equation}
%
See \cite{Raveri:2014cka,Kreisch:2017uet,Noller:2018wyv,Perenon:2019dpc,Arjona:2019rfn,Melville:2019wyy,SpurioMancini:2019rxy,Raveri:2019mxg,Brando:2019xbv,SolaPeracaula:2019zsl,SolaPeracaula:2020vpg,Joudaki:2020shz,Noller:2020afd,Traykova:2021hbr,Andrade:2023pws,Seraille:2024beb} for constraints on EFTofDE parameters from current data.  

\subsection{DESI forecast}\label{sec:DESIforecast}

\begin{figure}[h!]
    \centering
    \includegraphics[scale=0.7]{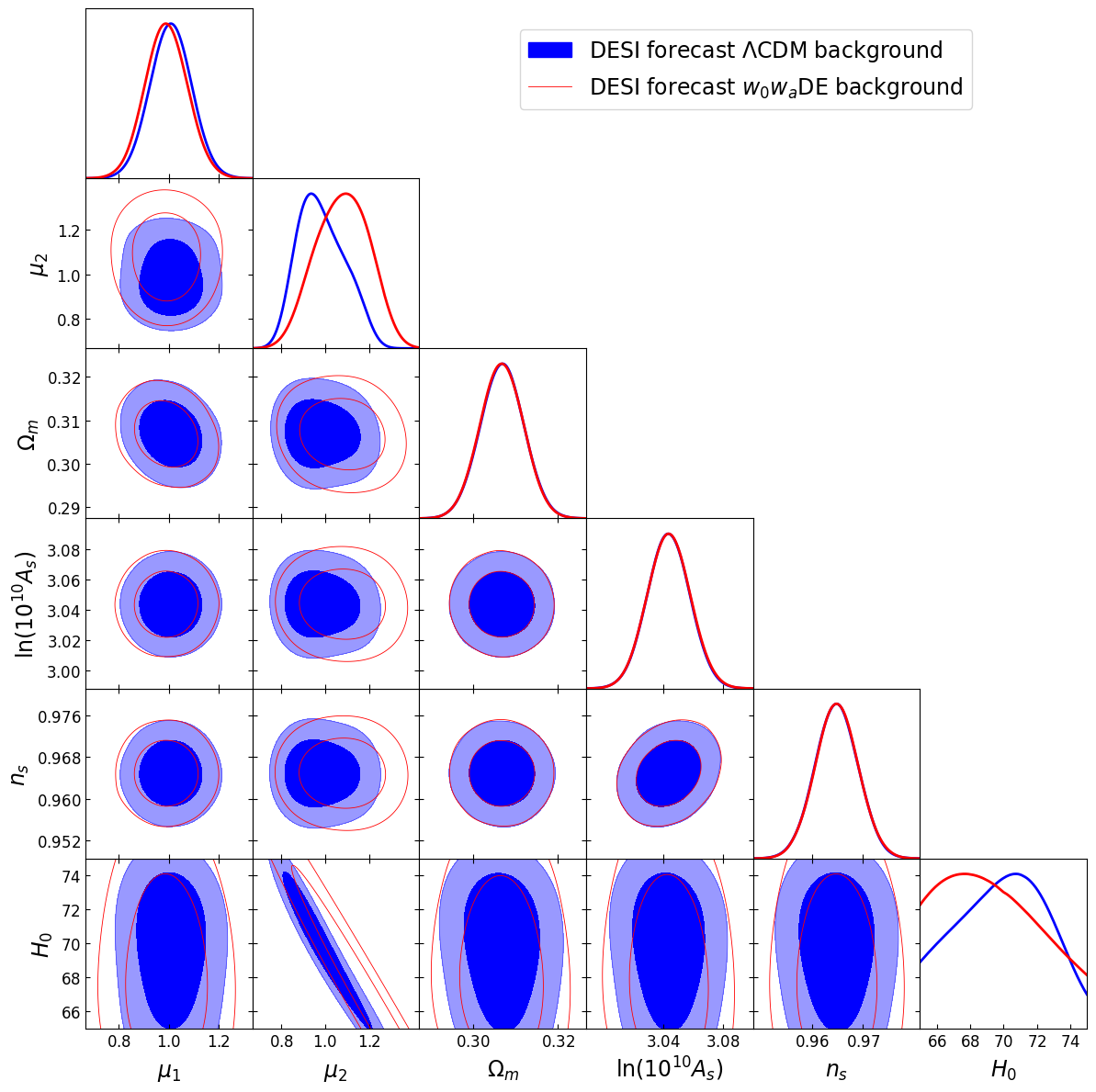}
    \caption{Contour plots and one-dimensional posterior distributions at 68\% and 95\% C.L. obtained from the DESI forecast. The latter is carried out assuming an underlying $\Lambda$CDM cosmology (in blue) and $w_0w_a$CDM cosmology (in red). $H_0$ is given in km/s/Mpc.}
    \label{fig:DESIforecast}
\end{figure}

The results of the forecast are displayed in Table \ref{tab:DESIforecast} and Fig. \ref{fig:DESIforecast}. We use the same priors on $A_s$, $n_s$, $\Omega_m$ and $H_0$ employed in Sec. \ref{sec:current_data}.  The covariance matrices obtained from the analysis of the mock data produced assuming a $\Lambda$CDM background with $\Omega_m=0.3085$ and a $w_0w_a$DE background with $(w_0,w_a,\Omega_m)=(-0.827,\,-0.75,\,0.3085)$ read, respectively, 

\begin{equation}\label{eq:cov_current_data} 
C[\mu_1, \mu_2,\Omega_m]=\begin{pmatrix}
6.74\cdot 10^{-3} & -2.03\cdot 10^{-4} & -8.35\cdot 10^{-5}\\
-2.03\cdot 10^{-4} & 1.19\cdot 10^{-2} & -5.21\cdot 10^{-5}\\
-8.35\cdot 10^{-5} &  -5.21\cdot 10^{-5}& 2.49\cdot 10^{-5}
\end{pmatrix}\,,
\end{equation}
and

\begin{equation}\label{eq:cov_current_data} 
C[\mu_1, \mu_2,\Omega_m]=\begin{pmatrix}
7.05\cdot 10^{-3} & -1.98\cdot 10^{-4} & -8.06\cdot 10^{-5}\\
-1.98\cdot 10^{-4} & 1.42\cdot 10^{-2} & -6.99\cdot 10^{-5}\\
-8.06\cdot 10^{-5} &  -6.99\cdot 10^{-5}& 2.47\cdot 10^{-5}
\end{pmatrix}\,.
\end{equation}
We do not observe significant differences in the results obtained with the two background cosmologies. Remarkably, our forecast predicts a substantial decrease in the uncertainties of both $\mu_1$ and $\mu_2$ compared to those obtained with current data, by approximately 40\% and 20\%, respectively, demonstrating the power of future DESI RSD data. The projected constraints on $c_B$ in the propto-Omega and Inv-Hubble-Squared models are displayed in the lower part of Table \ref{tab:DESIforecast}.

\begin{table}[h!]
\begin{center}
\begin{tabular}{| c | c |c |}
\multicolumn{1}{c}{Parameters} &  \multicolumn{1}{c}{$\Lambda$CDM mock data} &  \multicolumn{1}{c}{$w_0w_a$CDM mock data} 
\\\hline 
$\mu_1$ & $1.01\pm 0.08$ &  $0.99\pm 0.08$ 
\\\hline
$\mu_2$ & $0.98^{+0.09}_{-0.13}$ &  $1.08\pm 0.12$ 
\\\hline
$\Omega_m$ & $0.307\pm 0.005$ &  $0.307\pm 0.005$ 
\\\hline
$\ln(10^{10}A_s)$ & $3.044\pm 0.014$  & $3.044\pm 0.014$  
\\\hline
$n_s$ & $0.965\pm 0.004$ &  $0.965\pm 0.004$ 
\\\hline
$H_0$ [km/s/Mpc]& $70.0^{+3.0}_{-2.5}$ &  $69.4^{+1.4}_{-4.3}$ 
\\\hline
$\chi_{\rm min}^2$ & 0.06 & 0.10  
\\\hline\hline 
 \multirow{2}{*}{$c_B$ propto-Omega} & $<$0.71 (68\% C.L.) & $<$0.74 (68\% C.L.)
\\ & $<$1.35 (95\% C.L.) & $<$1.53 (95\% C.L.) \\\hline
$c_B$ Inv-Hubble-Squared & $<$0.57 (68\% C.L.) & $<$0.50 (68\% C.L.)
\\ & $<$1.06 (95\% C.L.) & $<$0.99 (95\% C.L.) \\\hline
 \end{tabular}
\end{center}
\caption{Mean values and corresponding uncertainties at $68\%$ C.L. obtained in the forecast with DESI. In the last row of the first block we report the minimum $\chi^2$, i.e., $\chi^2_{\rm min}$. We also present the 95\% C.L. uncertainties for the projected constraints on $c_B$. We consider the propto-Omega and Inv-Hubble-Squared models with $c_M=0$. See Sec. \ref{sec:DESIforecast} for details.}
\label{tab:DESIforecast}
\end{table}

\section{Conclusions}\label{sec:conclusions}
We applied the compression technique proposed in \cite{Denissenya:2017uuc} and studied its robustness and universality by applying it to three different models within the large class of EFTofDE. We considered three common parametrisations of the EFT functions employed in the literature. We modelled the gravitational coupling as a step function and introduce
2 or 3 parameters representing the step widths of the functions. We
demonstrated that this model could reproduce the growth of density fluctuations
with observational accuracy sufficient for DESI final year data. We provided a fitting formula for the gravitational coupling as a function of the EFTofDE parameter, which can be used to convert the constraints on the gravitational coupling to those on the EFT parameters. We showed that this compression technique could  reproduce the constraints on the EFT parameters in all models. In a model where the EFT parameter is proportional to the scale factor, the modification of gravity appears early and three bins are required. In other models where the EFT parameter is proportional to the dark energy density or inversely proportional to the Hubble function squred as inspired by the scaling solution, two bins are enough. 

The 2-3 parameter description offers a model-independent way to reveal the physical characteristics of the underlying theory required to explain the RSD data. We applied this technique to current RSD measurements and translated the constraints on the gravitational coupling to the EFT parameters. The current data is not sufficient to break the degeneracy between the second and third (the highest redshift) bin in the three parameter model, but the constraint on the first (the lowest redshift bin below $z=1$) is consistent between the two and three parameter model showing the suppression of the growth detecting $\mu_1 <1$ at the $\sim 2.6 - 2.8 \sigma$ confidence level. There results are aligned with those reported in previous works in the literature. Since the EFTofDE models that we considered in this paper predicts the enhancement of the growth, we get a tighter constraint on the EFT parameter than we expect. 

We also provided forecasts for DESI final year data using the predictions of $f \sigma_8$. The constraint on the gravitational constant and the EFT parameters required an assumption on the background expansion. We tested the effect of assuming the $\Lambda$CDM background on the constraints on the gravitational constant by creating synthetic data using the $w_0-w_a$ background and fitting the 2-parameter gravitational constant assuming the $\Lambda$CDM background. We used the bestfit $w_0-w_a$ form the recent DESI BAO measurement. The constraint in the lower redshift bin ($z<1$) is again robust. The constraint in the higher redshift bin is slightly biased but the input parameter is recovered well within $1 \sigma$. We also found that $\mu_2$ in the higher redshift bin degenerates with $H_0$ through $\sigma_8$. The expected uncertanties on $\mu_1$ and $\mu_2$ decrease by $\sim 40\%$ and $\sim 20\%$, respectively, compared to those obtained with current constraints, demonstrating a substantial improvement in the constraining power of DESI RSD data.

Although we focused on obtaining the constraints on $\mu$ using the compressed observable $f\sigma_{8}$ in this paper, our method can also be eventually applied to data on $f\sigma_{12}$ \cite{Sanchez:2020vvb,eBOSS:2021poy,Semenaite:2022unt}, in case they become available in the future. The 2-3 parameter description of $\mu$ can be also implemented in the EFTofLSS approach to obtain the constraints on $\mu_i$ directly. 
It is known that the EFTofLSS approach suffers from prior projection effects due to a large number of nuisance parameters.  Ref.~\cite{Piga:2022mge} tested a specific model, which gives a scale independent linear modification as in EFTofDE models considered in this paper and found a strong projection effect, i.e. even if the synthetic data is created by $\Lambda$CDM, the 1D marginalised constraint strongly favours non-zero modified gravity parameter after marginalising over nuisance parameters of EFTofLSS. We expect to see a similar effect for $\mu$ and this would become more severe with the increased number of parameters as they lead to more degeneracies among $\mu$ in different bins and the primordial amplitude. A fewer bins would alleviate this problem.  The 2-3 parameter description is ideal as we do not need to assume a specific time dependence of $\mu$, and it is general enough to cover a wide variety of models as shown in this paper and \cite{Denissenya:2017uuc}.  It remains to see whether the same binning of $\mu$ is enough for other LSS observables such as CMB (ISW, lensing) and weak lensing. This will be addressed in future work.


\section*{Acknowledgments}
This work was supported by JST SPRING, Grant Number JPMJSP2119 (YT). AGV is funded by “la Caixa” Foundation (ID 100010434) and the European Union's Horizon 2020 research and innovation programme under the Marie Sklodowska-Curie grant agreement No 847648, with fellowship code LCF/BQ/PI21/11830027. KK is supported by STFC grant ST/W001225/1. YT and AGV are grateful to the Institute of Cosmology and Gravitation of the University of Portsmouth for its kind hospitality during the first stages of this project.

\appendix

\section{How to implement $\mu(z)$ in CLASS}\label{sec:AppendixA}

In this Appendix we employ the notation of the seminal paper by Ma \& Bertschinger (Ma95, for short) \cite{Ma:1995ey}. \texttt{CLASS} \cite{Lesgourgues:2011re,Blas:2011rf} works with the following two perturbed Einstein equations in the Newtonian gauge: 

\begin{equation}\label{eq:psi}
\psi = \phi -12\pi G\frac{a^2}{k^2}(\rho+p)\sigma\,,
\end{equation}
\begin{equation}\label{eq:dotphi}
\phi^\prime=-\mathcal{H}\psi+4\pi G \frac{a^2}{k^2} (\rho+p)\theta\,,
\end{equation}
where the primes denote derivatives with respect to the conformal time and $\mathcal{H}=aH$. These equations correspond to Eqs. (23d) and (23b) of Ma95. They are implemented in the perturbation module of \texttt{CLASS}. We want to implement the parametrization 

\begin{equation}
\phi=\eta\psi\,,
\end{equation}
\begin{equation}\label{eq:PoissonMod}
k^2\psi = -4\pi Ga^2\mu\rho\Delta=-4\pi Ga^2\mu\rho\left(\delta+\frac{3\mathcal{H}\theta}{k^2}\right)\,,
\end{equation}
which allows for deviations from the standard model when $\mu\ne 1$ and/or $\eta\ne 1$. The following are the modified equations to be implemented in \texttt{CLASS}, 

\begin{equation}\label{eq:psi}
\psi = \frac{\phi}{\eta} -12\pi G\frac{a^2}{k^2}(\rho+p)\sigma\,,
\end{equation}
\begin{equation}\label{eq:dotphi}
\phi^\prime=\frac{\phi}{1+f}\left(\frac{\eta^\prime}{\eta}+\frac{\mu^\prime}{\mu}-\mathcal{H}\right)+\frac{f}{1+f}\left[-\mathcal{H}\psi+\theta_m\left(\frac{1}{3}-\frac{\mathcal{H}^\prime}{k^2}+\frac{\mathcal{H}^2}{k^2}\right)\right]\,,                                                
\end{equation}
with the function

\begin{equation}
f= 12\pi Ga^2\rho_m\frac{\mu\eta}{k^2}\,.
\end{equation}
We have checked that these equations coincide with those implemented in \texttt{MGCLASS II}, although we detected some typos in Eq. (4.10) of the associated paper \cite{Sakr:2021ylx}. This has motivated this appendix. 

In this work we are only considering the case $\eta=1$ ($\eta^\prime=0$) and the functions $\mu(a)$ provided in Eqs. \eqref{eq:muI} and \eqref{eq:muII}. Their derivatives are trivially computed as
\begin{equation}
    \mu^\prime=a\mathcal{H}\frac{d\mu}{da}\,.
\end{equation}

\bibliography{ref}

\end{document}